\documentclass{aa}

\usepackage[varg]{txfonts}
\usepackage{graphicx}
\usepackage{hyperref}
\usepackage{xcolor}
\usepackage{algorithm}
\usepackage{algorithmicx}
\usepackage{algpseudocode}
\usepackage{natbib}
\bibpunct{(}{)}{;}{a}{}{,}

\hypersetup{colorlinks=true,citecolor=blue,linkcolor=blue}

\newcommand{\vect}[1]{\mathit{\boldsymbol{#1}}}

\begin{document}

\title{
    A close-encounter method for simulating \\ the dynamics of planetesimals
    }

\author{
    Sebastian~Lorek\inst{1},
    Anders~Johansen\inst{1,2}
    }

\institute{
    Lund Observatory, 
    Department of Astronomy and Theoretical Physics, 
    Lund University,
    Box 43, 
    221 00 Lund,
    Sweden\\
    \email{lorek@astro.lu.se}
    \and
    Centre for Star and Planet Formation,
    Globe Institute, 
    University of Copenhagen,
    {\O}ster Voldgade 5–7, 
    DK-1350 Copenhagen, 
    Denmark
    }

\date{Received ; accepted }

\abstract{
	The dynamics of planetesimals plays an important role in planet formation, because their velocity distribution sets the growth rate to larger bodies. When planetesimals form in the gaseous environment of protoplanetary discs, their orbits are nearly circular and planar due to the effect of gas drag. However, mutual close encounters of the planetesimals increase eccentricities and inclinations until an equilibrium between stirring and damping is reached. After disc dissipation, there is no more gas that damps the motion and mutual close encounters as well as encounters with planets stir the orbits again. The high number of planetesimals in protoplanetary discs renders it difficult to simulate their dynamics by means of direct $N$-body simulations of planet formation. Therefore, we developed a novel method for the dynamical evolution of planetesimals that is based on following close encounters between planetesimal-mass bodies and gravitational stirring by planet-mass bodies. To separate the orbital motion from the close encounters, we employ a Hamiltonian splitting scheme as used in symplectic $N$-body integrators. Close encounters are identified using a cell algorithm with linear scaling in the number of bodies. A grouping algorithm is used to create small groups of interacting bodies which are integrated separately. Our method allows simulating a high number of planetesimals interacting through gravity and collisions with low computational cost. The typical computational time is of the order of minutes or hours, up to a few days for more complex simulations, as compared to several hours or even weeks for the same setup with full $N$-body. The dynamical evolution of the bodies is sufficiently well reproduced. This will make it possible to study the growth of planetesimals through collisions and pebble accretion coupled to their dynamics for a much higher number of bodies than previously accessible with full $N$-body simulations.
}


\maketitle

\section{Introduction}
\label{sec:introduction}
Planets grow in protoplanetary discs by the gradual accumulation of material, starting with micrometre-sized dust grains. Sticking collisions lead to the formation of millimetre-sized pebbles, because of growth barriers, such as radial drift, bouncing, and fragmentation, that prevent the bodies from growing to larger sizes \citep{Weidenschilling1977,Blum2008,Brauer2008,Guettler2010,Zsom2010,Birnstiel2012}. Therefore, other mechanisms are necessary to form planetesimals. The streaming instability is currently thought to be the main channel that leads to the formation of planetesimals which have sizes from typically a hundred kilometres up to about Ceres size \citep{Youdin2005,Johansen2007,Johansen2014,Carrera2015,Simon2016,Schaefer2017,Yang2017}. These planetesimals grow to protoplanets and the cores of giant planets through mutual collisions and the accretion of the remnant millimetre-sized pebbles \citep{Ormel2010,Lambrechts2012,Ormel2012,Ormel2017,Lambrechts2019,Liu2019}. Lastly, collisions between protoplanets and the accretion of gas from the protoplanetary disc by protoplanets more massive than $\sim10\,M_\oplus$ lead to the formation of terrestrial-like planets and gas giants \citep{Pollack1996,Raymond2009,Bitsch2019,Schulik2019}.

In the intermediate phase where planetesimals grow by collisions and pebble accretion, their dynamics plays a crucial role. The velocity distribution of planetesimals directly affects the accretion rate. However, the total number of planetesimals in protoplanetary discs is high and, in general, the gravitational interaction between all of them needs to be taken into account. Recent progress in the development and optimisation of $N$-body integrators which use an efficient splitting of the $N$-body Hamiltonian \citep{Wisdom1991,Duncan1998,Chambers1999,Rein2015}, have made it possible to perform dynamical studies of Solar System formation and evolution. However, when included, numerous planetesimal-sized bodies are typically treated either as non-interacting test particles which are only scattered and accreted by planets or as super particles which represent a subset of the planetesimals.

Because of the low density of planetesimal discs, stirring of eccentricity and inclination comes mainly from the cumulative effect of the scattering of two planetesimals that undergo close encounters in the gravitational field of the Sun. Therefore, a different approach to study planetesimal dynamics is to use statistical methods. Based on three-body calculations for the outcome of planetesimal scattering and averaging over an assumed velocity distribution, semi-analytic models \citep{Stewart1980,Stewart1988,Ida1990,Ohtsuki1999,Stewart2000,Ohtsuki2002} successfully reproduce $N$-body simulations \citep{Ida1992a,Ida1992b,Ida1993,Palmer1993,Aarseth1993} and provide a way to account for planetesimal dynamics in toy models of planet formation \citep[e.g.][]{Chambers2006a,Chambers2006b,Ormel2012}.

A third approach to model planetesimal dynamics comes from the fact that the dominant contribution to orbital changes are because of close encounters between two planetesimals \citep{Henon1986,Petit1986,Weidenschilling1989,Hasegawa1990,Ida1990}. Therefore, as in the statistical approach, planetesimal dynamics is simulated as the cumulative effect of close encounters. This type of methods were already used by \citet{Cox1980,Lecar1986,Beauge1990}. \citet{Cox1980} evolved planetesimals along Keplerian orbits as long as their distance was larger than the Tisserand sphere of influence. Inside this sphere, the close encounter was resolved by assuming hyperbolic trajectories for both planetesimals. A different approach was chosen by \citet{Lecar1986}. Instead of resolving the close encounter only when inside the sphere of influence, they divided the orbital plane into radial and azimuthal zones to include the gravitational perturbations of the nearest neighbours, given in terms of the grid, when integrating the equations of motion of the planetesimals. \citet{Lecar1986} pointed out that the method of \citet{Cox1980} misses out the close encounters that dominate the stirring because the Tisserand sphere of influence is smaller than the relevant encounter distance they derived, which is of the size of the Hill radius.

Here, we chose a hybrid approach combining the ideas of symplectic $N$-body methods \citep[e.g.][]{Chambers1999} and close encounters \citep{Cox1980,Lecar1986}. Instead of a statistical treatment of planetesimal dynamics, we develop a method that efficiently evolves a high number of planetesimals on their orbits taking mutual gravitational interaction into account. This method has the advantage that we can simulate many bodies in reasonable computational time, in contrast to conventional $N$-body methods, and do not rely on averaging over distribution functions, as done in the statistical methods. We achieve this by separating the dynamics of the planetesimals into unperturbed Keplerian motion and two-body scattering in the Solar gravitational field and by using an efficient method for the detection of close encounters. To make the method even more versatile and applicable to studies of planet formation, we treat planets in the conventional $N$-body sense.

We outline the method in Sect.~\ref{sec:method}, where we start by introducing the concept of symplectic $N$-body integrators and continue with a technical description of the different components of our method. In Sect.~\ref{sec:test}, we present the results of different test cases, performance, and scaling tests. Limitations are discussed in Sect.~\ref{sec:discussion} before we summarise and conclude in Sect.~\ref{sec:summary}. Algorithms, our prescription for pebble accretion, gas drag, and migration can be found in the Appendix Sects.~\ref{sec:algorithms} to \ref{sec:gasdragandmigration}.

\section{Method}
\label{sec:method}
We construct our close-encounter method based on the theory of symplectic integrators for the gravitational $N$-body problem. Symplectic integrators have become popular for this because they use the Hamiltonian to construct efficient and accurate numerical schemes. Symplectic integrators preserve the Hamiltonian structure by conserving the phase space volume and by solving a Hamiltonian that is close to the original one, which means energy is conserved \citep{Yoshida1993,Hernandez2017}.

\subsection{Theory of symplectic integrators}
The Hamiltonian of the gravitational $N$-body problem is the sum of kinetic and potential energy,
\begin{equation}
\mathcal{H}=\sum_{i=1}^{N}\frac{\vect{p}_i^2}{2m_i}-\sum_{i=1}^{N}\sum_{j=i+1}^{N}\frac{Gm_im_j}{|\vect{x}_i-\vect{x}_j|},
\end{equation}
where $\vect{p}_i=m_i\vect{v}_i$ and $\vect{x}_i$ are the momentum and the position of body $i$, and the equations of motion are given by Hamilton's equations
\begin{equation}
	\dot{\vect{x}_i} = \frac{\partial \mathcal{H}}{\partial \vect{p}_i}, \qquad
	\dot{\vect{p}_i} = -\frac{\partial \mathcal{H}}{\partial \vect{x}_i}, \qquad
	i\in(1,\dots,N).
\end{equation}

The time derivative of any quantity $f$ that is a function of the $\vect{x}_i$ and $\vect{p}_i$, can then be written as
\begin{equation}
	\dot{f} = \sum_{i=1}^{N}\left(\frac{\partial f}{\partial \vect{x}_i}\frac{\partial\mathcal{H}}{\partial \vect{p}_i}-\frac{\partial f}{\partial \vect{p}_i}\frac{\partial\mathcal{H}}{\partial \vect{x}_i}\right)=\hat{D}f, \label{eq:eomoperator}
\end{equation}
where $\hat{D}$ is a differential operator \citep{Hanslmeier1984,Chambers1999}. We can now formally integrate Eq~\ref{eq:eomoperator} to obtain the general solution of $f$ at time $t+\Delta t$
\begin{equation}
	f(t+\Delta t)=\mathrm{e}^{\hat{D}\Delta t}\,f(t)
\end{equation}
\citep[e.g.][]{Hanslmeier1984,Yoshida1993,Hernandez2017,Wisdom2018,Tamayo2019}. Therefore, the solution of Eq.~\ref{eq:eomoperator} is formally given by an ideal operator $\mathcal{D}=\mathrm{e}^{\hat{D}\Delta t}$ which depends on the Hamiltonian $\mathcal{H}$ and advances $f$ by one time step $\Delta t$. The exponential of an operator is defined as 
\begin{equation}
	\mathrm{e}^{\hat{D}\Delta t}=\sum_{k=0}^{\infty}\frac{1}{k!}\left(\hat{D}\Delta t\right)^k=1+\Delta t\hat{D}+\frac{1}{2}\Delta t^2\hat{D}^2+\dots \label{eq:operatorexp}
\end{equation}
\citep{Hanslmeier1984}. However, $\mathcal{D}$ has no closed-form solution in general. Therefore, we make use of splitting the Hamiltonian into parts each of which can be solved easily on its own and applied such as to approximate the true solution of the problem \citep[e.g.][]{Wisdom1991,Duncan1998,Chambers1999}.

For example, if we split the Hamiltonian into two parts, such that $\mathcal{H}=\mathcal{H}_\mathrm{A}+\mathcal{H}_\mathrm{B}$, we would get two operators $\hat{A}$ and $\hat{B}$
\begin{equation}
	\hat{A}=\sum_{i=1}^{N}\left(\frac{\partial}{\partial \vect{x}_i}\frac{\partial\mathcal{H}_\mathrm{A}}{\partial \vect{p}_i}-\frac{\partial}{\partial \vect{p}_i}\frac{\partial\mathcal{H}_\mathrm{A}}{\partial \vect{x}_i}\right),
\end{equation}
and a similar expression for $\hat{B}$, such that $\mathcal{D}=\mathrm{e}^{(\hat{A}+\hat{B})\Delta t}$. However, $\hat{A}$ and $\hat{B}$ do not necessarily commute, which means that the relation $[\hat{A},\hat{B}]=\hat{A}\hat{B}-\hat{B}\hat{A}\neq0$ in general. With this in mind, one can easily show that by expanding the exponential expression
$\mathrm{e}^{(\hat{A}+\hat{B})\Delta t}=\mathrm{e}^{\hat{A}\Delta t}\mathrm{e}^{\hat{B}\Delta t}$ is only true to first order in $\Delta t$. Higher order solutions can be constructed in the following way
\begin{equation}
	\mathrm{e}^{(\hat{A}+\hat{B})\Delta t}=\prod_{i=1}^{k}\mathrm{e}^{c_i\hat{A}\Delta t}\mathrm{e}^{d_i\hat{B}\Delta t}+\mathcal{O}(\Delta t^{n+1})
\end{equation}	
for a suitable set of coefficients $(c_i,d_i)$ with $i\in(1,\dots,k)$ \citep{Yoshida1993}. For example, a second-order accurate scheme is obtained by setting $c_1=1/2=c_2$, $d_1=1$, and $d_2=0$ which results in
\begin{equation}
	\mathrm{e}^{(\hat{A}+\hat{B})\Delta t}=\mathrm{e}^{\frac{1}{2}\hat{A}\Delta t}\mathrm{e}^{\hat{B}\Delta t}\mathrm{e}^{\frac{1}{2}\hat{A}\Delta t}+\mathcal{O}(\Delta t^3) \label{eq:secondorder}
\end{equation}
\citep{Yoshida1993}.

Splitting the Hamiltonian in kinetic and potential energy is often used for the general $N$-body problem which allows us to construct the second-order accurate leap-frog algorithm. For the Solar System, however, where the Sun is the dominating body forcing the other bodies on Keplerian orbits, a different splitting is more beneficial. Using so-called democratic-heliocentric coordinates, which means heliocentric position and barycentric momentum, the Hamiltonian can be transformed to
\begin{equation}
	\mathcal{H} = \mathcal{H}_\mathrm{A} + \mathcal{H}_\mathrm{B} + \mathcal{H}_\mathrm{C} \label{eq:hfull},
\end{equation}
with
\begin{subequations}
\begin{align}
	\mathcal{H}_\mathrm{A} &= \sum_{i=1}^{N}\left(\frac{\vect{p}_i^2}{2m_i}-\frac{Gm_im_\odot}{|\vect{r}_i|}\right), \label{eq:hkep} \\
	\mathcal{H}_\mathrm{B} &= -\sum_{i=1}^{N}\sum_{j=i+1}^{N}\frac{Gm_im_j}{|\vect{r}_i-\vect{r}_j|}, \label{eq:hint} \\
	\mathcal{H}_\mathrm{C} &= \frac{1}{2m_\odot}\left(\sum_{i=1}^{N}\vect{p}_i\right)^2 \label{eq:hsun}.
\end{align}
\end{subequations}
Here, $\vect{r}_i=\vect{x}_i-\vect{x_\odot}$ is the heliocentric position of body $i$ and $\vect{p}_i$ is its barycentric momentum. $\mathcal{H}_\mathrm{A}$ is the Hamiltonian of the Kepler problem. $\mathcal{H}_\mathrm{B}$ takes the interactions between the bodies, excluding the Sun, into account. $\mathcal{H}_\mathrm{C}$ is the kinetic energy of the Sun. Each of the Hamiltonians can be easily solved in the absence of the others. Efficient methods exist for solving the Kepler problem \citep{Danby1992,Rein2015,Wisdom2015}, the interaction part reduces to calculating forces between the bodies, and $\mathcal{H}_\mathrm{C}$ does not depend on the position, which makes it easy to solve as well. Therefore, we can use this splitting to construct a second-order accurate scheme in the form of Eq.~\ref{eq:secondorder} as
\begin{equation}
	\mathrm{e}^{(\hat{A}+\hat{B}+\hat{C})\Delta t}=\mathrm{e}^{\frac{1}{2}\hat{B}\Delta t}\mathrm{e}^{\frac{1}{2}\hat{C}\Delta t}\mathrm{e}^{\hat{A}\Delta t}\mathrm{e}^{\frac{1}{2}\hat{C}\Delta t}\mathrm{e}^{\frac{1}{2}\hat{B}\Delta t}+\mathrm{O}(\Delta t^3) \label{eq:nbodyscheme}.
\end{equation}
to integrate the motion of a body $i$ \citep{Chambers1999}, that is for $f=(\vect{r}_i,\vect{p}_i)$. The first step is a kick for half of a time step owing to the interaction with the other bodies. Then, the body drifts from the change in kinetic energy of the Sun. Next, the body moves on a Keplerian orbit for $\Delta t$, followed by another drift and a kick.

\subsection{Keplerian motion}
\label{sec:unperturbedkeplerianmotion}
Without mutual gravitational perturbation, bodies move on Keplerian orbits. The equivalent one-body Hamiltonian of Keplerian motion for body $i$ is
\begin{equation}
	\mathcal{H}_\mathrm{A}^i=\frac{\vect{p}_i}{2m_i}-\frac{Gm_im_\odot}{|\vect{r}_i|},
\end{equation} 
We use the method described in \citet{Mikkola1999,Rein2015} which uses the Gauss $f$- and $g$-functions to solve for it.

The Gauss $f$- and $g$-function formalism is advantageous for efficiently solving the Kepler problem because it avoids computationally expensive transformations to orbital elements and problems arising with circular orbits \citep{Rein2015}.

The key idea is that the position $\vect{r}_i$ and velocity $\vect{v}_i=\vect{p}_i/m_i$ of a body at time $t=t_0+\Delta t$ can be expressed in terms of the initial values at time $t_0$, $\vect{r}_{i,0}$ and $\vect{v}_{i,0}$, as
\begin{align}
	\vect{r}_i&=f\vect{r}_{i,0}+g\vect{v}_{i,0}, \\
	\vect{v}_i&=\dot{f}\vect{r}_{i,0}+\dot{g}\vect{v}_{i,0},
\end{align}
where $f$ and $g$ are scalar functions which only depend on time
and $\dot{f}$ and $\dot{g}$ are the time derivatives of these functions. A detailed description of how to obtain $f$ and $g$, their time derivatives, and how to implement and use them to calculate the orbit can be found, for instance, in \citet{Danby1992,Mikkola1999,Rein2015,Wisdom2015} and references therein.

\subsection{Interactions between the bodies}
\label{sec:closeencounterdetection}
The integration scheme of Eq.~\ref{eq:nbodyscheme} constructed from the Hamiltonian splitting of Eq.~\ref{eq:hfull} can be used as long as bodies are well separated. Because the Keplerian part dominates in that case, the error for integrating one time step is $\sim\mathcal{O}(\epsilon\Delta t^3)$ with $\epsilon$ being the mass ratio of the bodies with respect to the Sun \citep{Chambers1999}. However, when two bodies have a close encounter, the interaction term becomes comparable to $\mathcal{H}_\mathrm{A}$ and the error per time step increases which prevents us from obtaining accurate results. To avoid this problem, hybrid methods were developed which make sure that $\mathcal{H}_\mathrm{B}$ remains small compared to the Keplerian part all the time. The key idea is to move the interacting bodies from  $\mathcal{H}_\mathrm{B}$ to $\mathcal{H}_\mathrm{A}$ and to integrate those bodies numerically with a higher-order method instead of using a Kepler solver. We can move bodies between $\mathcal{H}_\mathrm{A}$ and $\mathcal{H}_\mathrm{B}$ by replacing Eqs.~\ref{eq:hkep} and \ref{eq:hint} with
\begin{align}
	\mathcal{H}_\mathrm{A} &= \sum_{i=1}^{N}\left(\frac{\vect{p}_i^2}{2m_i}-\frac{Gm_im_\odot}{|\vect{r}_i|}\right) \nonumber \\ &+ \sum_{i=1}^{N}\sum_{j=i+1}^{N}\frac{Gm_im_j}{|\vect{r}_i-\vect{r}_j|}\left[1-K(r_{ij})\right], \\
	\mathcal{H}_\mathrm{B} &= -\sum_{i=1}^{N}\sum_{j=i+1}^{N}\frac{Gm_im_j}{|\vect{r}_i-\vect{r}_j|}K(r_{ij})
\end{align}
where $K$ is a function of the mutual distance of two planetesimals, $r_{ij}=|\vect{r}_i-\vect{r}_j|$, which acts as a switch to move terms between the two Hamiltonians \citep[e.g.][]{Duncan1998,Chambers1999,Rein2019a}. The corresponding operators then become
\begin{align}
	\hat{A} &= \sum_{i=1}^{N}\left\{\frac{\partial}{\partial \vect{r}_i}\frac{\vect{p}_i}{m_i}-\frac{\partial}{\partial\vect{p}_i}Gm_im_\odot\frac{\vect{r}_i}{|\vect{r}_i|^3}\right. \nonumber \\
	&\qquad\quad\left.-\frac{\partial}{\partial\vect{p}_i}Gm_i\sum_{j\neq i}^{N}m_j\frac{\vect{r}_i-\vect{r}_j}{|\vect{r}_i-\vect{r}_j|^3}\left[1-L(r_{ij})\right]\right\}, \label{eq:kepleroperator} \\
	\hat{B} &= \sum_{i=1}^{N}\left\{\frac{\partial}{\partial\vect{p}_i}Gm_i\sum_{j\neq i}^{N}m_j\frac{\vect{r}_i-\vect{r}_j}{|\vect{r}_i-\vect{r}_j|^3}L(r_{ij})\right\},
\end{align}
where we defined a new function $L=K-r_{ij}\partial K/\partial r$ \citep{Rein2019a}. We chose $L$ to be a Heaviside step function which is defined such that
\begin{equation}
	L =
	\begin{cases}
		0 & r_{ij}<d_\mathrm{ce}, \\
		1 & r_{ij}>d_\mathrm{ce},
	\end{cases}
\end{equation}
where $d_\mathrm{ce}$ is the critical distance between bodies for classifying it as a close encounter (see Sect.~\ref{sec:distanceforcloseencounter}). It is now easy to see that without a close encounter, $\hat{A}$ reduces to the operator for the Kepler problem, which can be easily solved using a Kepler solver, and $\hat{B}$ adds small perturbations. For a close encounter, on the other hand, $\hat{A}$ contains an additional term and we resort to numerically integrating a few-body problem using a Bulirsch-Stoer algorithm \citep{Press1992}.

So far, we have only summarised the basic ideas of hybrid-symplectic integrators for the planetary $N$-body problem. However, a bottleneck of simulating a high number of bodies is their mutual interaction (i.e. $\mathcal{H}_\mathrm{B}$). Because forces between every body pair must be calculated, which is an $N^2$ calculation, simulating many bodies would result in unfeasibly long computation times. We have approached this problem by dividing the bodies into two classes, namely planets and planetesimals, and by using an efficient close-encounter detection method.

\paragraph{Planets}
\label{sec:planets}
For planets, we cannot ignore the long-range forces and therefore treat them in full $N$-body fashion. We solve the full Hamiltonian, which means that we not only resolve close encounters, but also take the distant perturbations into account. Therefore, our method is equivalent to an hybrid-symplectic integrator, such as \texttt{mercury} \citep{Chambers1999}.

\paragraph{Planetesimals}
We take planetesimals as low-mass bodies for which we chose to ignore the long-range interaction and only take close encounters into account. This approach is justified because of the low total mass of the planetesimals compared to the Sun and the fact that significant orbital changes only occur if the approach is very close, typically within a few times the Hill radius \citep{Petit1986,Greenberg1991}. The strategy is thus to evolve planetesimals on their Keplerian orbit if there is no close encounter, as described in the previous section, and to evolve the few-body problem of the Sun plus nearby planetesimals otherwise. Because planetesimals grow by mutual collisions and pebble accretion, they naturally evolve into planets. To account for this, we promote planetesimals to planets, when they reach a minimum mass, which we take as $10^{-2}\,M_\oplus$.

\subsubsection{Distance for close encounter}
\label{sec:distanceforcloseencounter}
If the distance $r_{ij}$ between two bodies $i$ and $j$ is less than a critical distance $d_\mathrm{ce}$, they have a close encounter. This critical distance is typically a few times the Hill radius
\begin{equation}
	R_\mathrm{h}=r\left(\frac{m}{3m_\odot}\right)^{1/3}.
\end{equation}
Here, $r$ is the radial heliocentric distance and $m$ is the mass of the body. Within a sphere of radius $R_\mathrm{h}$, the gravity of the body dominates over the solar gravity for the motion of a test particle in the co-rotating frame. As a consequence, the mutual gravitational interaction between the two bodies results in strong scattering and significant changes of their respective orbits \citep{Henon1986,Petit1986,Ida1990}. Distant encounters, where $r_{ij}\gg R_\mathrm{h}$, only lead to weak scattering with small orbital changes \citep{Weidenschilling1989,Hasegawa1990}. The same applies for encounters with $r_{ij}\ll R_\mathrm{h}$ which result in horseshoe orbits. In these type of encounters, eccentricity and inclination hardly change \citep{Henon1986,Hasegawa1990}. Therefore, only if $r_{ij}\sim R_\mathrm{h}$, the encountering body can enter the Hill sphere of the other object and will be strongly scattered.

In our close-encounter method, $d_\mathrm{ce}$ in units of the Hill radius, is a parameter which is set for each body individually. Numerical simulations of planetesimal encounters showed that orbital changes are strong if $0.8\,R_\mathrm{h}\la r_{ij} \la2.5\,R_\mathrm{h}$ \citep{Petit1986,Greenberg1991}. Therefore, a typical value to use in our method would be $d_\mathrm{ce}\sim3\,R_\mathrm{h}$. We do not use a lower value which would result in missing close encounters. On the other hand, a larger value catches also more distant encounters and, thus, we typically chose a value of $d_\mathrm{ce}\sim3\dots5\,R_\mathrm{h}$ which reproduces the dynamics of the planetesimals to a high degree as we show below (see Sect.~\ref{sec:test}).

\subsubsection{Cell-list approach}
A computationally efficient way to detect close encounters is with a cell list, an approach that is frequently used in molecular dynamics simulations \citep{Frenkel2002} and dust particle collisions \citep{Johansen2012}. A grid divides the simulation domain into $n_\mathrm{cell}$ individual cells. The grid size is chosen in a way that the cell dimensions are larger than the typical distance associated with a close encounter, that is $\sim d_\mathrm{ce}$. For each cell, we assign a body as head of the list and link it to the next one in this cell, which creates a linked list of nearby bodies (see Appendix~\ref{sec:algorithms} Alg.~\ref{alg:celllist} for the algorithm in pseudo-code).

To identify a close encounter, we loop through the cell list. The criterion for a close encounter is that the distance between two bodies is less than $d_\mathrm{ce}$. Bodies residing close to the cell boundaries potentially undergo a close encounter with an object from a neighbouring cell. Therefore, looping over the neighbour cells is necessary to account for these cases as well. There are at most $27$ cells to loop over. To check for close encounters, the orbits of the bodies are approximated as straight lines connecting their initial positions and the new positions they would have if advanced along their unperturbed Keplerian orbit. If the minimum separation between those lines is less than $d_\mathrm{ce}$, the encounter is classified as a close encounter and the body pair is added to a list (see Appendix~\ref{sec:algorithms} Alg.~\ref{alg:detectce} for the algorithm in pseudo-code) which is used afterwards to integrate the encounter (see Sect.~\ref{sec:closeencounterintegration}).

Because creating the cell list scales linearly with the number of bodies and because the average number of bodies per grid cell is lower than the total number in the simulation, the close-encounter detection is faster than a simple loop over each body pair, which scales as $N^2$, especially for a high number of bodies.

\subsubsection{Grouping into close-encounter groups}
\label{sec:groupingintocloseencountergroups}
If more than two bodies have a close encounter, which may occur for high surface density of planetesimals or for large $d_\mathrm{ce}$, the bodies need to be grouped together. To do so, we follow the grouping algorithm of \citet{Grimm2014} (see Appendix~\ref{sec:algorithms} Alg.~\ref{alg:groupingce} for the algorithm in pseudo-code).

\subsubsection{Close-encounter integration}
\label{sec:closeencounterintegration}
Having a list of close encounters, we numerically integrate the motion of these bodies. To do so, we use the Bulirsch-Stoer (BS) method \citep{Press1992} as implemented, for example, also in \texttt{mercury} \citep{Chambers1999}. The basic idea of the BS method is to integrate the motion of the body for one time step $\Delta t$ by sub-dividing $\Delta t$ into smaller bits and integrate those with a modified midpoint rule until the results change by less than a given tolerance parameter.

The gravitational acceleration acting on body $i$ according to Eq.~\ref{eq:kepleroperator} is
\begin{equation}
	\vect{a}_i=-Gm_\odot\frac{\vect{r}_i}{|\vect{r}_i|^3}-\sum_{j\neq i}^{N_\mathrm{ce}}Gm_j\frac{\vect{r}_j-\vect{r}_i}{(r_{ij}^2+b^2)^{3/2}}.
\end{equation}
The first term is the gravity of the Sun and the second term is the acceleration from the other bodies. The summation is over all bodies $N_\mathrm{ce}$ that are involved in the close encounter. We use a gravitational softening parameter $b$ to avoid high acceleration in very close encounters. The softening parameter is set to be the sum of the physical radii of planetesimals $i$ and $j$. This choice for the softening is reasonable because the two planetesimals would undergo a physical collision if $r_{ij}\le b$. If additional forces are included, they are simply added to $\vect{a}_i$.

We provide two different ways to integrate the close encounters. A sequential approach which uses the list of close-encounter pairs and a group approach which uses the grouping into close-encounter groups.

\paragraph{Sequential approach}
The sequential approach is appropriate if planetesimals have at most one encounter per time step. We therefore have $N_\mathrm{ce}=1$ which means that we solve the three-body problem of the Sun and two other bodies. To avoid correlations in rare cases when a body might encounter more than one other body, we randomise the close-encounter list.

\paragraph{Group approach}
The group approach is useful if the bodies encounters more than one other body per time step. In that case, we have to treat the encounter group as a small $N$-body system with $N=N_\mathrm{ce}>1$ bodies. For large groups with $N_\mathrm{ce}\gg1$, the group approach becomes comparable to a standard hybrid-symplectic scheme. Therefore, large groups should be avoided to obtain the full potential of the close-encounter method.

\subsection{Grid size and time step}
\label{sec:gridsizeandtimestep}
Using a cell list requires a grid. We chose to construct the grid in cylindrical coordinates $(r,\phi,z)$, which is especially useful for simulating rings of planetesimals. The grid cells need to be large enough to include the close-encounter region of a body. This sets the minimum size for the grid cells to be $\Delta r>d_\mathrm{ce}$, $\Delta \phi>d_\mathrm{ce}/r$, and $\Delta z>d_\mathrm{ce}$. If the grid cell was smaller, bodies would potentially undergo close encounters with bodies outside the neighbouring cells. On the other hand, the grid cells need to be small enough, such that the average number of planetesimals per grid cell is $\ll N$. Otherwise, the cell list approach loses its scaling-advantage over a conventional $N^2$-pair search.

Using a cell list also imposes a constraint on the maximum time step that we can use in the simulation. As pointed out before, only objects in the current and neighbouring cells are checked for close encounters. Therefore, bodies should not move over more than one grid cell per time step to avoid missing close encounters with bodies outside the neighbouring cells. To adjust the time step, we calculate the crossing times in each direction for all bodies. The current allowed time step is then the minimum of these crossing times
\begin{align}
	\Delta t=\min_{i}\left(
		\frac{\Delta r}{|v_{r,i}},
		\frac{\Delta \phi}{|v_{\phi,i}|},
		\frac{\Delta z}{|v_{z,i}|}
		\right),
\end{align}
where $i$ runs over all bodies and the radial, azimuthal, and vertical velocities of the bodies are $v_{r,i}$, $v_{\phi,i}$, and $v_{z,i}$, respectively. Because the time step is not constant throughout the simulation, our method is not symplectic.

\subsection{Additional forces}
\label{sec:additionalforces}
To be able to include gas drag or migration, we allow for additional forces in our method. We do so by adding the acceleration in the interaction part that comes from $\mathcal{H}_\mathrm{B}$ \citep{Chambers1999,Tamayo2019}.

\section{Results}
\label{sec:test}
We test the close-encounter method by simulating the viscous stirring of a ring of planetesimals and compare the results to $N$-body simulations done with \texttt{mercury} \citep{Chambers1999} and to the semi-analytical toy model of \citet{Ohtsuki2002}.

We use the same setup as in \citet{Ohtsuki2002} to test the viscous stirring of planetesimals. A ring of $1000$ planetesimals with surface density of $10\,\mathrm{g}\,\mathrm{cm}^{-2}$ is located at $1\,\mathrm{au}$. For a single planetesimal mass, we use a mass of $m=10^{24}\,\mathrm{g}$. For a bimodal mass distribution, we split the $1000$ planetesimals into $800$ with mass $10^{24}\,\mathrm{g}$ (component 1) and $200$ with mass $4\times10^{24}\,\mathrm{g}$ (component 2). The initial orbits have eccentricities and inclinations following Rayleigh-distributions with $e_\mathrm{rms}=10^{-4}$ and $i_\mathrm{rms}=5\times10^{-5}$. The phase angles, argument of perihelion $\omega$, longitude of ascending node $\Omega$, and the mean anomaly $M$ are drawn from a uniform distribution in $(0,2\pi)$. We then integrate the planetesimals for a total time of $10^4\,\mathrm{yr}$ \citep[][only integrated for $10^3\,\mathrm{yr}$]{Ohtsuki2002}. In \texttt{mercury}, we use the hybrid-symplectic method, a time step of $8\,\mathrm{days}$, and a Bulirsch-Stoer tolerance of $10^{-12}$. For both, our model and the comparison runs with \texttt{mercury}, we show averages over five runs with varying random seed for setting up the initial conditions.

\subsection{Equal-mass planetesimals}
\label{sec:equalmassplanetesimals}
\begin{figure}
	\resizebox{\hsize}{!}{
		\includegraphics{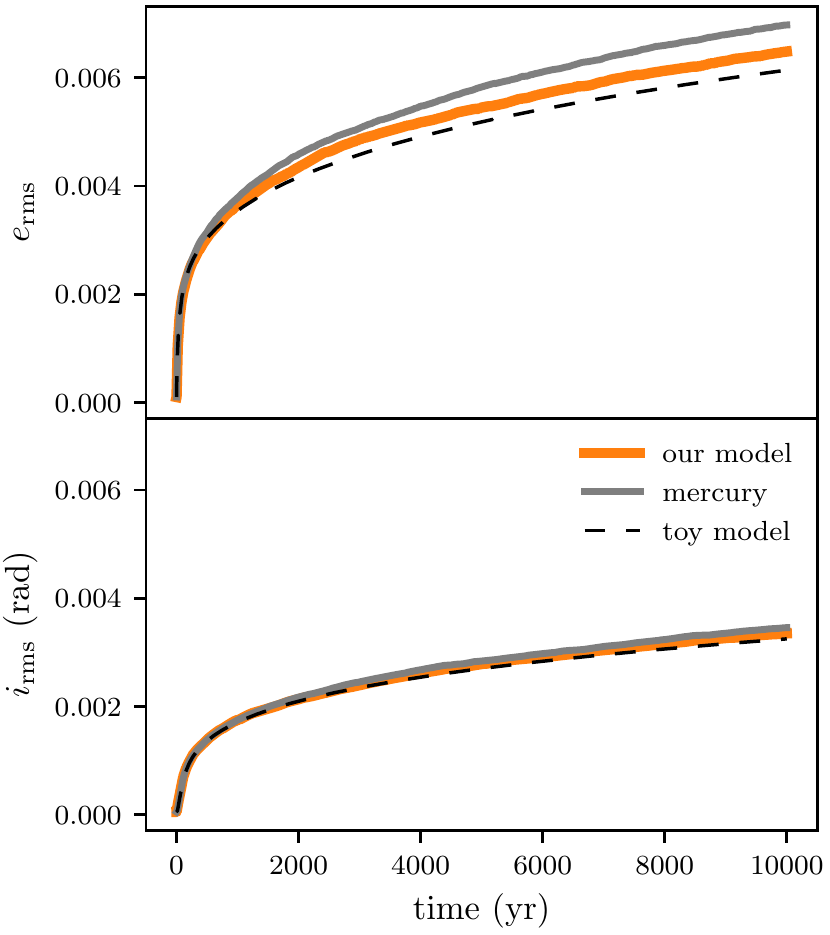}}
	\caption{Viscous stirring for equal-mass planetesimals. The figure shows root mean-square eccentricity (\textit{top panel}) and inclination (\textit{bottom panel}) of a ring of equal-mass planetesimals with mass $1\times10^{24}\,\mathrm{g}$. The ring is centred at $1\,\mathrm{au}$ and the surface density is $10\,\mathrm{g}\,\mathrm{cm}^{-2}$. In both panels, we show the result of our model, the close-encounter approach (orange), a full $N$-body simulation with \texttt{mercury} (grey), and the semi-analytic toy model of \citet{Ohtsuki2002} (black-dashed).}
	\label{fig:paper1_testcase1comp}
\end{figure}
Figure~\ref{fig:paper1_testcase1comp} shows the viscous stirring of the equal-mass planetesimals. For the close-encounter distance, we used $10$ Hill radii in our model. Every encounter with a minimum distance less than that is integrated with the Bulirsch-Stoer part of our code. In this test case, we use the group approach. Our method reproduces the results of the full $N$-body simulation and the toy model sufficiently well, showing that the close-encounter approach works well for this particular case. We have also tested the sequential approach for this specific test case using $5$ Hill radii for the close-encounter distance and found no noticeable difference other than a speed-up of $\sim1.5$.

\subsection{Unequal-mass planetesimals}
\label{sec:unequalmassplanetesimals}
\begin{figure}
	\resizebox{\hsize}{!}{
		\includegraphics{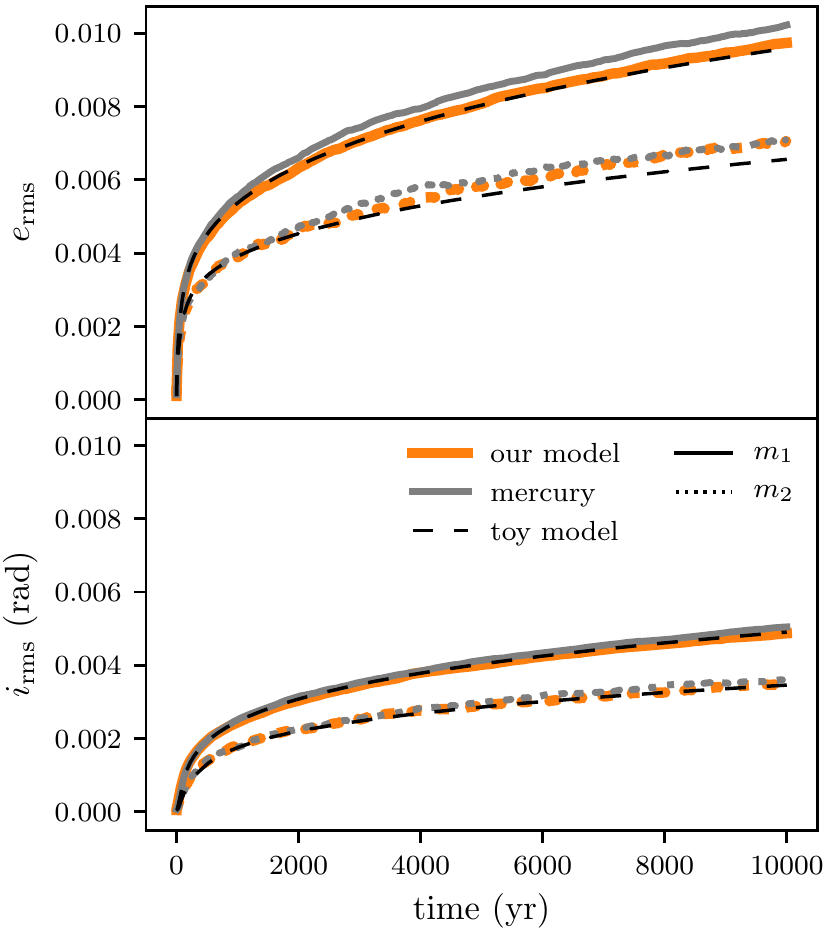}}
	\caption{Viscous stirring for bimodal mass distribution. The figure shows root mean-square eccentricity (\textit{top panel}) and inclination (\textit{bottom panel}) of a ring of planetesimals with bimodal mass distribution. There are $800$ planetesimals with mass $10^{24}\,\mathrm{g}$ (component 1, $m_1$) and $200$ with mass $4\times10^{24}\,\mathrm{g}$ (component 2, $m_2$). The ring is centred at $1\,\mathrm{au}$ and the surface density is $10\,\mathrm{g}\,\mathrm{cm}^{-2}$. In both panels, we show the result of our model (orange), a full $N$-body simulation with \texttt{mercury} (grey), and the semi-analytic model of \citet{Ohtsuki2002} (black-dashed). Component 1 and 2 are indicated with solid and dotted lines, respectively.}
	\label{fig:paper1_testcase2comp}
\end{figure}
Figure~\ref{fig:paper1_testcase2comp} shows the viscous stirring and dynamical friction of the two planetesimal populations. Our model, the close-encounter method, matches the \texttt{mercury} and toy-model results. Because the gravitational kicks on the lower mass bodies (component 1) from the more massive planetesimals (component 2) are stronger than vice versa, $e_\mathrm{rms}$ and $i_\mathrm{rms}$ are on average higher for component 1. This effect is called dynamical friction which drives the system towards energy equipartition and a mass-dependent velocity dispersion. However, equipartition is not exactly reached because viscous stirring also increases $e_\mathrm{rms}$ and $i_\mathrm{rms}$ of both components.

\subsection{Embedded planet}
\label{sec:embeddedplanet}
\begin{figure}
	\resizebox{\hsize}{!}{
		\includegraphics{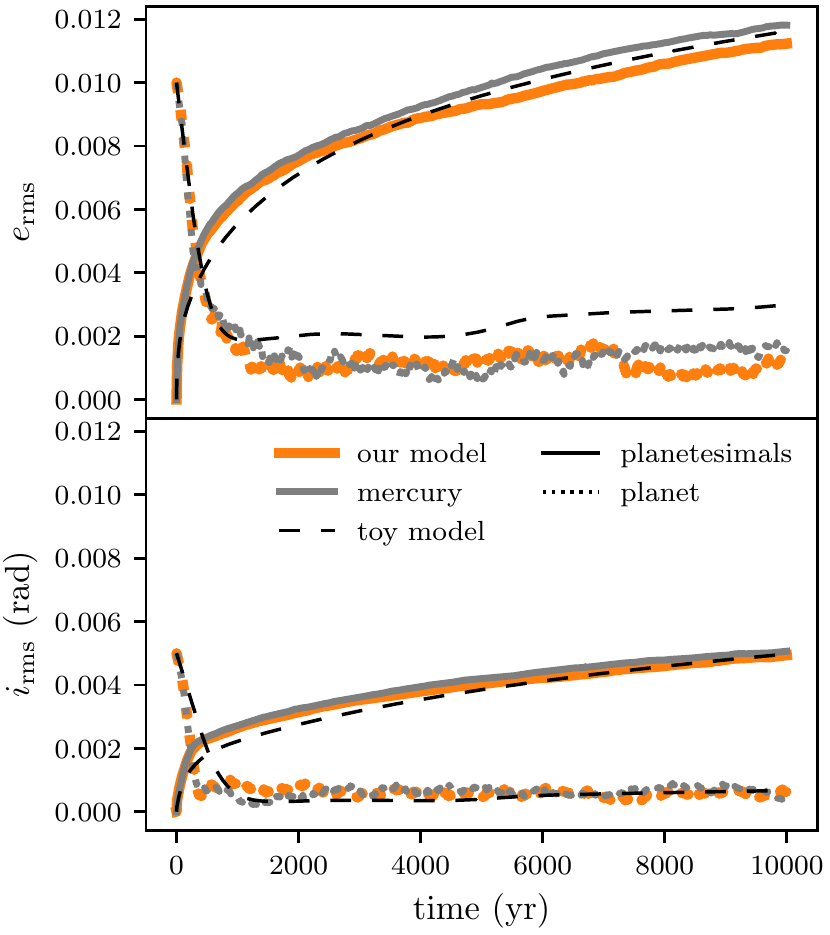}}
	\caption{Viscous stirring and dynamical friction for a planet embedded in a ring of planetesimals. The figure shows root mean-square eccentricity (\textit{top panel}) and inclination (\textit{bottom panel}) of a ring of planetesimals with an embedded planet. There are $1000$ planetesimals with mass $10^{24}\,\mathrm{g}$ and $1$ planet with mass $10^{26}\,\mathrm{g}$. The ring is centred at $1\,\mathrm{au}$ and the surface density is $10\,\mathrm{g}\,\mathrm{cm}^{-2}$. The planet is initially at $1\,\mathrm{au}$. In both panels, we show the result of our model (orange), a full $N$-body simulation with \texttt{mercury} (grey), and the semi-analytic model of \citet{Ohtsuki2002} (black-dashed). Planetesimals and the planet are indicated with solid and dotted lines, respectively.}
	\label{fig:paper1_testcaseembp}
\end{figure}
In this test case shown in Fig.~\ref{fig:paper1_testcaseembp}, we simulate the viscous stirring of a ring of planetesimals with an embedded planet. To see the effects of viscous stirring and dynamical friction, we initialise the planet with $e=10^{-2}$ and $i=5\times10^{-3}$, which are $1000$ times larger than the rms-values of the planetesimals, which we initialised here with $e_\mathrm{rms}=10^{-5}$ and $i_\mathrm{rms}=5\times10^{-6}$. We see that the orbit of the planet is rapidly damped towards very low eccentricity and inclination. This is the effect of dynamical friction through the swarm of planetesimals. On the other hand, the planetesimals are excited into highly eccentric and inclined orbits. Also here, our method matches the results of $N$-body simulations and the toy model.

\subsection{Dependence on the close-encounter distance}
\label{sec:dependenceonthecloseencounterdistance}
\begin{figure}
	\resizebox{\hsize}{!}{
		\includegraphics{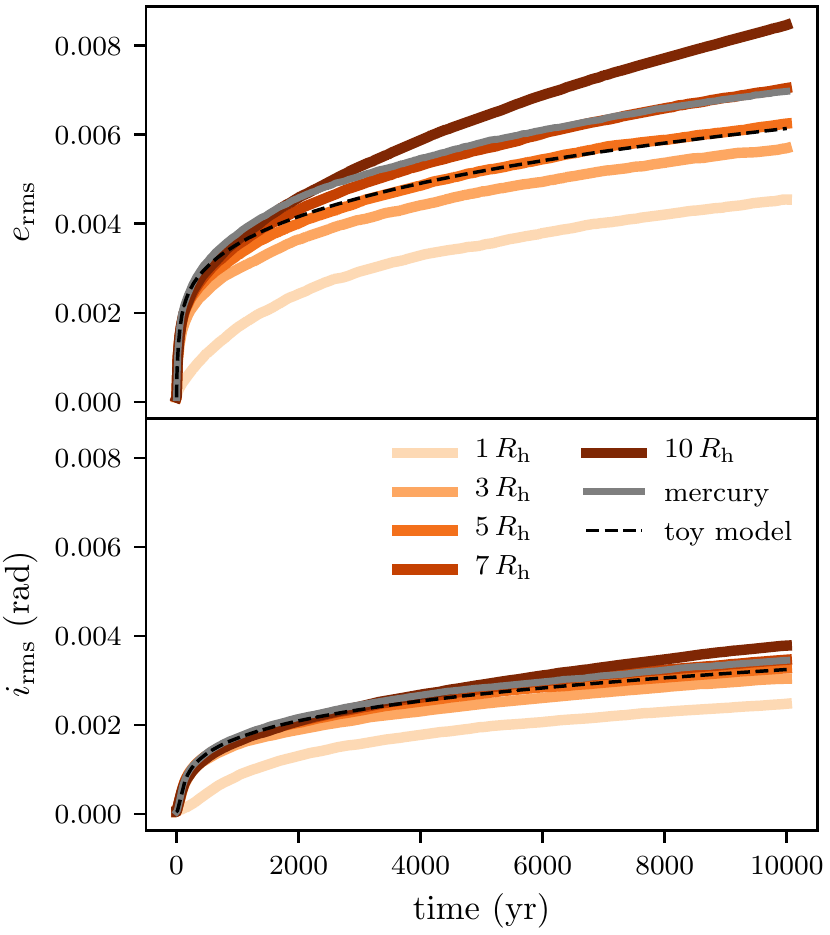}}
	\caption{Viscous stirring of equal-mass planetesimals for increasing close-encounter distance using the sequential approach. The figure shows root mean-square eccentricity (\textit{top panel}) and inclination (\textit{bottom panel}) for the same setup as for the equal-mass planetesimal run. In both panels, we show the result of our model for different values of $d_\mathrm{ce}$ (coloured lines), a full $N$-body simulation with \texttt{mercury} (grey), and the semi-analytic model of \citet{Ohtsuki2002} (black-dashed).}
	\label{fig:paper1_dcemax}
\end{figure}
The critical distance for close encounters $d_\mathrm{ce}$ controls the contribution of distant encounters to the stirring of the planetesimals. The stirring increases for larger $d_\mathrm{ce}$ because more distant encounters are included. We investigate how the choice of $d_\mathrm{ce}$ affects our method. Therefore, we run the equal-mass planetesimal setup for different values of $d_\mathrm{ce}$, ranging from $1\,R_\mathrm{h}$ to $10\,R_\mathrm{h}$.

Figure~\ref{fig:paper1_dcemax} shows the results when using the sequential approach in our method. For $d_\mathrm{ce}=1\,R_\mathrm{h}$, we miss many close encounters and therefore get to little stirring of the planetesimals. Increasing it to $d_\mathrm{ce}\sim5-7\,R_\mathrm{h}$ produces a close match with the toy model and \texttt{mercury}. Increasing $d_\mathrm{ce}$ more results in increased stirring at times $t\ga3000\,\mathrm{yr}$. The reason is that the sequential approach breaks down. Because $d_\mathrm{ce}$ is large, the close-encounter regions of planetesimals overlap and they encounter more than one other body per time step. In this type of situation, integrating all close encounters sequentially gives the wrong results.

\begin{figure}
	\resizebox{\hsize}{!}{
		\includegraphics{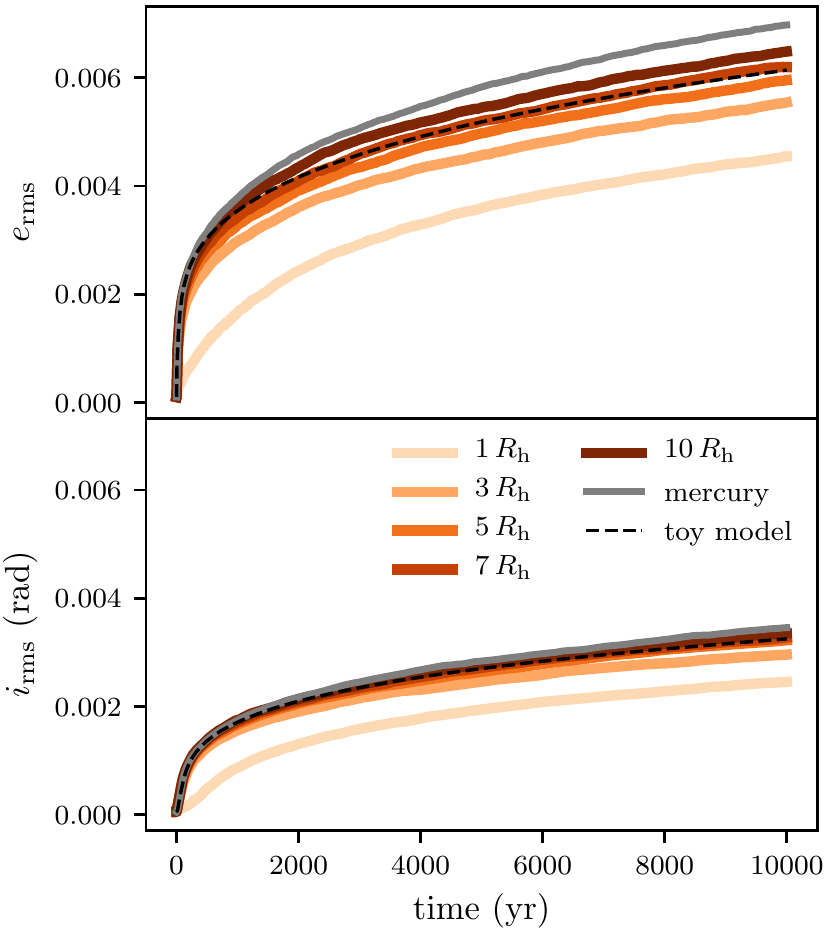}}
	\caption{Viscous stirring of equal-mass planetesimals for increasing close-encounter distance using the group approach. The figure shows root mean-square eccentricity (\textit{top panel}) and inclination (\textit{bottom panel}) for the same setup as for the equal-mass planetesimal run. In both panels, we show the result of our model for different values of $d_\mathrm{ce}$ (coloured lines), a full $N$-body simulation with \texttt{mercury} (grey), and the semi-analytic model of \citet{Ohtsuki2002} (black-dashed).}
	\label{fig:paper1_dcemax_group}
\end{figure}
Figure~\ref{fig:paper1_dcemax_group} shows the same experiment of increasing the close-encounter distance, but this time we used the group approach of our method. The results are similar for $d_\mathrm{ce}=1\,R_\mathrm{h}$, for which we miss too many encounters, but the solution converges towards the toy model and \texttt{mercury} for larger values. Therefore, the sequential approach is safe to use as long as there is only one encounter per planetesimal per time step. We will discuss this limitation more in Sect.~\ref{sec:discussion}.

\subsection{Computational time and scaling}
\label{sec:computationaltime}
\begin{figure}
	\resizebox{\hsize}{!}{
		\includegraphics{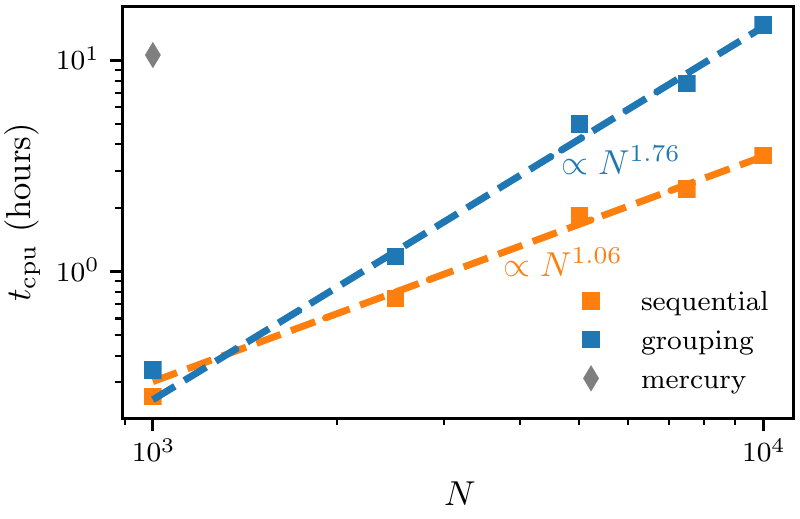}}
	\caption{Scaling of computational time with number of bodies. The setup is the same as in Sect.~\ref{sec:equalmassplanetesimals}. The surface density is kept constant at $10\,\mathrm{g}\,\mathrm{cm}^{-2}$. The number of planetesimals increases from $1000$ to $10\,000$. We show the total computational time to integrate $10^4\,\mathrm{yr}$ with the sequential approach (sequential; orange), the group approach (grouping; blue), and, for comparison, also with \texttt{mercury} (grey diamond). We also fitted a power-law to find the scaling with $N$ for our model (blue and orange lines, respectively).}
	\label{fig:paper1_scaling}
\end{figure}
We are aiming at developing a computationally fast method. Therefore, we look at the computational time and the scaling of our close-encounter method with number of bodies $N$ for the setup of a ring of equal-mass planetesimals. Figure~\ref{fig:paper1_scaling} shows the result. We measured the computational time for the sequential approach (sequential; orange), the group approach (grouping; blue), and, for comparison, a full $N$-body run with $1000$ bodies with \texttt{mercury}.

We can see the advantage of our method in Fig.~\ref{fig:paper1_scaling}. Simulating the dynamical evolution of $1000$ equal-mass planetesimals for $10^4\,\mathrm{yr}$ takes about $15$ minutes with the close-encounter method. The same setup with full $N$-body with \texttt{mercury} requires $10$ hours, and more for higher number of bodies. However, in $10$ hours, the close-encounter method -- both approaches, sequential and group -- can integrate the dynamical evolution of $10\,000$ bodies. The close-encounter method is hence faster without losing the ability to obtain correct results. Full $N$-body methods typically scale with the number of bodies as $N^2$. Our sequential approach scales linearly, which is expected because $N^2$ operations are avoided. Setting up the cell list is an operation of order $N$ and the integration of close encounters reduces to a sequence of three-body integrations. The group approach does not scale linearly with $N$, but better than quadratic as $\propto N^{1.8}$. The execution time is much faster than for conventional $N$-body methods in both cases. This allows simulating higher numbers of bodies than with full $N$-body methods.

\subsection{Energy and angular-momentum conservation}
\label{sec:energyandangularmomentumconservation}
\begin{figure}
	\resizebox{\hsize}{!}{
		\includegraphics{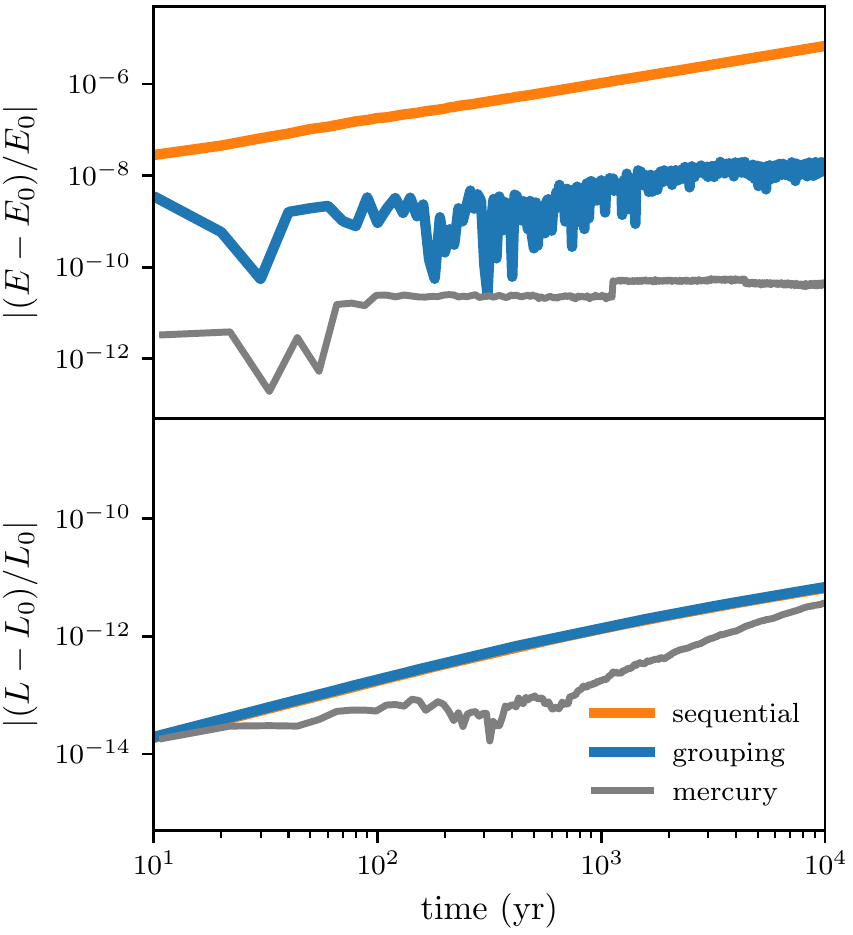}}
	\caption{Energy and angular momentum. The figure shows the time evolution of the relative change of energy (\textit{top panel}) and angular momentum (\textit{bottom panel}) for equal-mass planetesimals. We show the results obtained with the sequential approach (sequential; orange) and the group approach (grouping; blue) of our method and compare it with \texttt{mercury} (grey).}
	\label{fig:paper1_1comp_energy}
\end{figure}
We look at the energy and angular-momentum conservation of our close-encounter method by studying the relative change of energy and angular momentum with time and compare it to a full $N$-body simulation with \texttt{mercury}. We again use the setup for the equal-mass planetesimal ring. The relative change of energy $E$ is $|(E-E_0)/E_0|$ (and likewise for the angular momentum $L$), where $E_0$ is the total energy at the beginning and $E$ is the energy at time $t$.

Figure~\ref{fig:paper1_1comp_energy} shows the results. Energy and angular momentum show a steady increase over the simulation time of $10^4\,\mathrm{yr}$. We first look at the angular momentum. Regardless of the approach (sequential or group), the angular-momentum change is the same. Compared to \texttt{mercury}, we achieve the same accuracy. The situation looks slightly different for the energy. The hybrid-symplectic integrator \texttt{mercury} conserves energy. This is because \texttt{mercury} is a symplectic method and solves a Hamiltonian which is close to the true $N$-body Hamiltonian of the system at any time \citep{Chambers1999}. Our close-encounter method, however, has two significant differences. Firstly, we ignore the interaction terms for planetesimals unless they undergo a close encounter and only for planets, our method is comparable to \texttt{mercury}. Therefore, each time there is a close encounter between planetesimals, our method changes the Hamiltonian (that is the energy of the system) by adding a small perturbation owing to the interaction. Secondly, our method does not use a constant time step which breaks the symplecticity. These errors accumulate over time, as seen in Fig.~\ref{fig:paper1_1comp_energy}, which results in the steady increase in $|(E-E_0)/E_0|$.

\subsection{Collisions}
\label{sec:collisions}
\begin{figure*}
	\resizebox{\hsize}{!}{
		\includegraphics{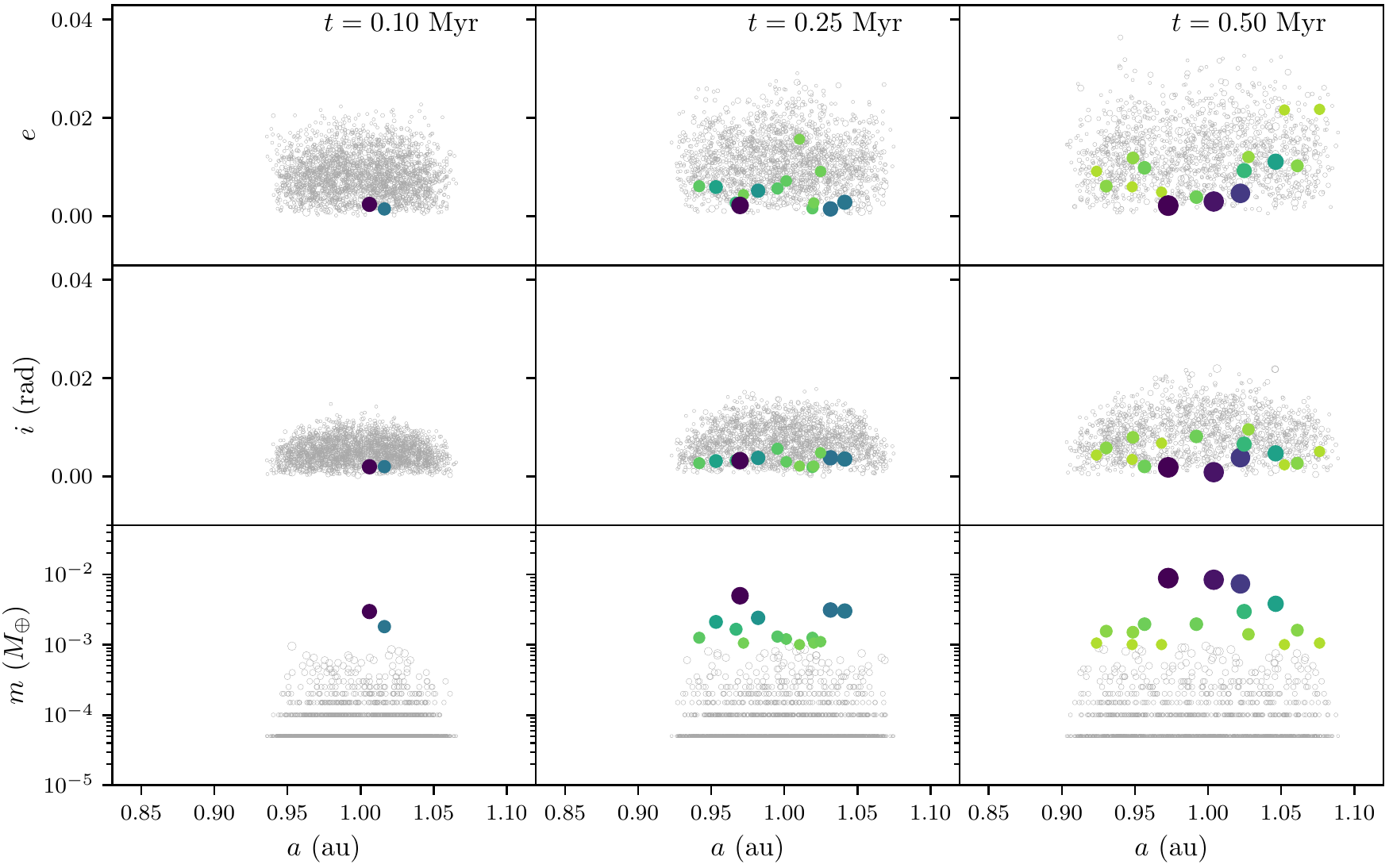}}
	\caption{Collisional growth of planetesimals at $1\,\mathrm{au}$. The Figure shows scatter plots for eccentricity (\textit{first row}), inclination (\textit{second row}), and mass (\textit{third row}). $4000$ planetesimals are located at $1\,\mathrm{au}$ in a ring of width $0.085\,\mathrm{au}$. The three columns show snapshots at different times. We highlight planetesimals that grow more massive than $10^{-3}\,M_\oplus$ as coloured filled circles. All other bodies are grey open circles. The colour and size of the symbols scale with the instantaneous mass.}
	\label{fig:paper1_testcasecoll}
\end{figure*}
We apply the close-encounter method to study the collisional growth of planetesimals located in a ring at $1\,\mathrm{au}$. The setup is similar to the one studied in \citet{Kokubo1998,Kokubo2000}. $4000$ planetesimals with mass $3\times10^{23}\,\mathrm{g}$ are placed at $1\,\mathrm{au}$ in a ring of width $0.085\,\mathrm{au}$. Initial eccentricities and inclinations are Rayleigh-distributed with $e_\mathrm{rms}=2i_\mathrm{rms}=2\times10^{-3}$. Gas drag is included as described in Sect.~\ref{sec:gasdrag}. Bodies that reach a mass of $10^{-2}\,M_\oplus$ are promoted to planets. We integrate for $0.5\,\mathrm{Myr}$. Figure~\ref{fig:paper1_testcasecoll} shows snapshots of eccentricity, inclination, and mass as a function of semi-major axis at three different times.

\begin{figure}
	\resizebox{\hsize}{!}{
		\includegraphics{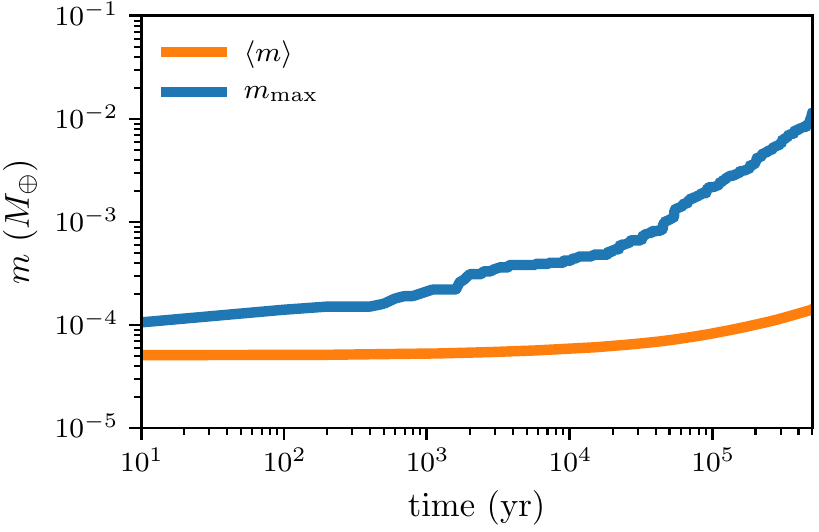}}
	\caption{Mean and maximum mass of planetesimals. We show the mean (orange line) and maximum (blue line) mass of the planetesimals as a function of time. Both lines are averages of five runs with different random realisations of the initial conditions.}
	\label{fig:paper1_testcasecollmeanmax}
\end{figure}
Starting with equal masses, collisions result in the formation of runaway bodies within $\sim5\times10^4\,\mathrm{yr}$. In the later evolution, these bodies grow fastest and separate from the bulk of the planetesimals. This can be seen in the bottom panels of Fig.~\ref{fig:paper1_testcasecoll}. Figure~\ref{fig:paper1_testcasecollmeanmax} shows the mean and maximum mass of the system. There, the runaway growth can be seen even better. The mean mass increases by about a factor of two. The maximum mass, however, increases more rapidly with time and separates from the mean mass, as it is characteristic for runaway growth. By the end of the simulation at $0.5\,\mathrm{Myr}$, the maximum mass increased by a factor of $\sim100$ whereas the mean mass increased by a factor of $\sim2$. This is consistent with \citet{Kokubo1998,Kokubo2000}. The few massive bodies furthermore start to stir the surrounding planetesimals, which is visible in the $a$-$e$ and $a$-$i$ plots (Fig.~\ref{fig:paper1_testcasecoll}). The lower mass planetesimals are excited to eccentricities and inclinations of $\sim0.02$. The more massive bodies, on the other hand, retain low eccentricities and inclinations because of dynamical friction. At $0.5\,\mathrm{Myr}$, a total number of $1484$ planetesimals ($m<10^{-2}\,M_\oplus$) and $2$ protoplanets ($m>10^{-2}\,M_\oplus$) remain. Simulating $0.5\,\mathrm{Myr}$ of evolution took about $2.3\,\mathrm{days}$. In comparison, with \texttt{mercury} it took us about $3$ weeks to simulate the same setup for $0.1\,\mathrm{Myr}$.

\subsection{Pebble accretion}
\label{sec:pebbleaccretion}
As another test application, we apply the close-encounter method to planet formation with collisions and pebble accretion in the giant planet region. Therefore, we distribute $6$ embryos of mass $10^{-2}\,M_\oplus$ and $444$ planetesimals of mass $10^{-4}\,M_\oplus$ in a ring between $5\,\mathrm{au}$ and $10\,\mathrm{au}$. A total flux of pebbles of $100\,M_\oplus\,\mathrm{Myr}^{-1}$ enters the system from the outer disc and is potentially accreted by the bodies. The pebbles have a constant Stokes number of $10^{-2}$ and the ratio of the pebble to the gas scale-height is $0.1$. We use a prescription for pebble accretion that is based on the work by \citet{Ormel2010,Lambrechts2012,Johansen2017,Ormel2017,Lambrechts2019}. Details can be found in the Appendix Sec.~\ref{sec:pebbleaccretion}. As bodies grow in mass, they the start to migrate. We therefore implemented an additional force for type-I migration following \citet{Cresswell2008} (see Appendix~\ref{sec:type1migration}). The gas disc is a Minimum Mass Solar Nebula \citep[][see Appendix~\ref{sec:gasdrag}]{Weidenschilling1977b,Hayashi1981}. We want to emphasize that this is only a demonstration of the capability of the close-encounter method without the goal of providing a quantitative study of pebble accretion.

\begin{figure*}
	\resizebox{\hsize}{!}{
		\includegraphics{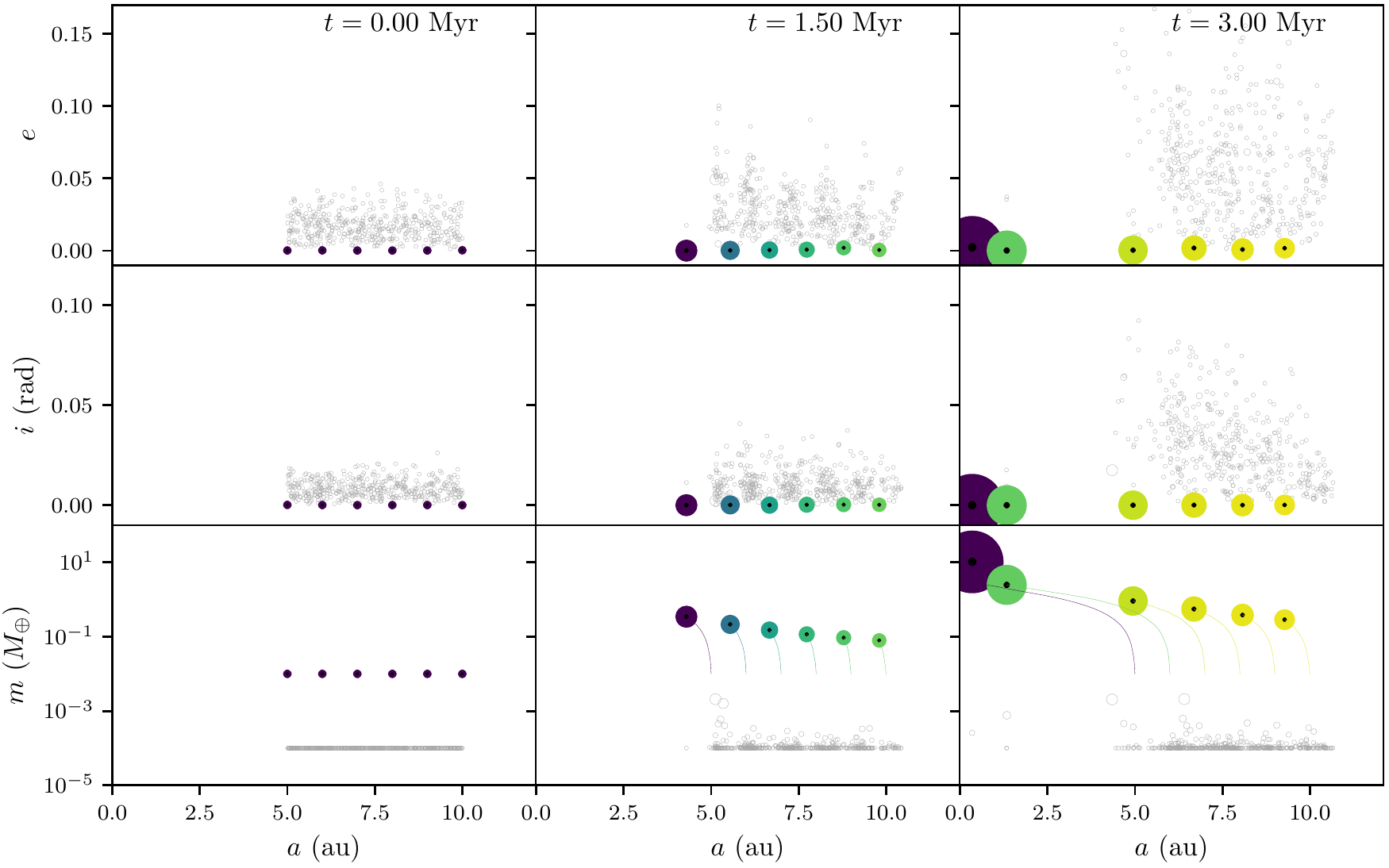}}
	\caption{Pebble accretion and collisions in the giant planet region. The Figure shows scatter plots for eccentricity (\textit{first row}), inclination (\textit{second row}), and mass (\textit{third row}). A total mass of $0.1\,M_\oplus$ is distributed amongst $450$ planetesimals between $5$ au and $10$ au. The three columns show snapshots at $0$ Myr, $1.5$ Myr, and at the final time of $3$ Myr, respectively. Bodies that are more massive than $10^{-2}\,M_\oplus$ are shown as coloured filled circles, their growth tracks are shown by the coloured lines, and the centre of the circle is indicated by a black dot. All other bodies are grey open circles. The colour and size of the symbols scale with the instantaneous mass.}
	\label{fig:paper1_testcasepa}
\end{figure*}
\begin{figure}
	\resizebox{\hsize}{!}{
		\includegraphics{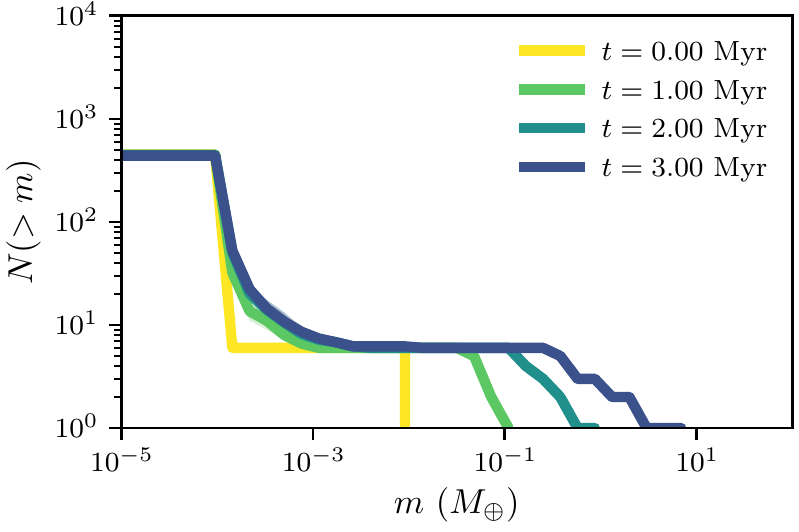}}
	\caption{Cumulative number of bodies for pebble accretion in the giant planet region. The figure shows $N(>m)$, the cumulative number of bodies with mass higher than $m$ for the initial distribution and for the distribution at times $0,\mathrm{Myr}$, $1\,\mathrm{Myr}$, $2\,\mathrm{Myr}$, and $3\,\mathrm{Myr}$. Each line is the average of five runs with randomised initial configuration of the bodies.}
	\label{fig:paper1_testcasepam}
\end{figure}

Figure~\ref{fig:paper1_testcasepa} shows snapshots of eccentricity, inclination, and mass, at times $0\,\mathrm{Myr}$, $1.5\,\mathrm{Myr}$, and the final outcome after $3\,\mathrm{Myr}$. The $6$ embryos grow efficiently by pebble accretion to a few Earth masses within $3\,\mathrm{Myr}$ (see Fig.~\ref{fig:paper1_testcasepam}). Collisions with planetesimals do not play a significant role. As the embryos grow bigger, they start to migrate inward by type-I migration. The most massive body reaches pebble isolation mass of $\sim10\,M_\oplus$ and stops migrating at the inner edge of the disc, which is located here at $0.3\,\mathrm{au}$. We do not form cold gas giants in our simulations. The assumed pebble flux of $100\,M_\oplus\,\mathrm{Myr}^{-1}$ corresponds to a nominal gas accretion-rate of the disc of $\sim3\times10^{-8}\,M_\odot\,\mathrm{yr}^{-1}$ \citep{Hartmann2016} for a dust-to-gas ratio of $0.01$ and a nominal metallicity. Slightly higher accretion rates and metallicities are needed to form cold gas giants within the disc lifetime \citep{Johansen2019a,Bitsch2020}.

\begin{figure}
	\resizebox{\hsize}{!}{
		\includegraphics{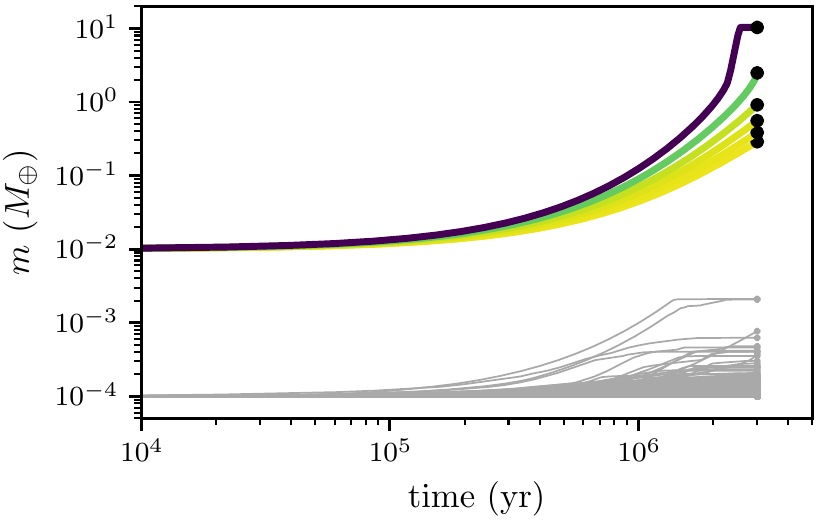}}
	\caption{Mass evolution of planetesimals for collisions and pebble accretion in the giant planet region. Each line shows the growth history of a planetesimal or embryo. Bodies that grow more massive than $10^{-2}\,M_\oplus$ are shown as coloured lines. The colour scales with mass and is the same as in Fig.~\ref{fig:paper1_testcasepa}. The grey lines show all other bodies. We indicate the endpoint of the simulation for each body with a circle. Because the bodies do not grow significantly before $10^{4}\,\mathrm{yr}$, we only show the later times.}
	\label{fig:paper1_testcasepamt}
\end{figure}

Pebble accretion proceeds in a runaway fashion, as seen in Fig.~\ref{fig:paper1_testcasepamt}. The embryos grow very fast, but the the majority of the planetesimals remain at low masses. The exponential growth comes from the fact that the embryos grow in the 3D regime \citep{Lambrechts2019}.

Figure~\ref{fig:paper1_testcasepa} also shows that the massive embryos efficiently stir the surrounding planetesimals which renders collisional growth inefficient. The combined effect of eccentricity and inclination damping due to the gas and dynamical friction with the lower mass planetesimals keeps the embryos on orbits with low eccentricities and inclinations throughout the simulation.

One simulation took about $3$ days to complete $3\,\mathrm{Myr}$ of evolution. This is considerably longer than the runs with only collisions. However, with pebble accretion more bodies potentially reach the threshold mass of $10^{-2}\,M_\oplus$ and need to be considered planets in our close-encounter method, which is computationally more expensive. Additionally, type-I migration reduces the maximum time step allowed in the simulation because of the grid which is necessary for creating the cell list, which requires more single integration steps.

\section{Discussion}
\label{sec:discussion}
As shown in Fig.~\ref{fig:paper1_scaling}, the sequential close-encounter approach scales almost linearly with the number of bodies. The group approach scales as $N^{1.8}$, which is only slightly better than quadratically. However, for both methods, the computational time is significantly lower than for \texttt{mercury}. Therefore, it is possible to simulate a high number of planetesimals in reasonable times. The speed-up is not only because we ignore the long-range interactions for planetesimals, and thus the $N^2$ process of calculating the forces between all planetesimals, but also because we chose a more efficient detection method for close encounters. The test cases, which were compared to $N$-body simulations with \texttt{mercury} and semi-analytic toy models, show that our approach sufficiently reproduces the dynamics of the planetesimals. When applied to more physical cases, that is including planets, collisions, and pebble accretion, our method becomes a faster alternative to conventional $N$-body methods.

The sequential treatment of close encounters increases the speed of the method significantly; however, one has to carefully check the validity of this approximation. The sequential approach breaks down when planetesimals have more than one encounter per time step. This can occur either in a very dense system or when the close-encounter regions of many planetesimals overlap. Because the physical sizes of planetesimals are small, the latter situation is much more likely. The group approach and conventional $N$-body codes do not suffer from this limitation, but integrating the encounters sequentially will give wrong results.

We will now derive an approximate criterion for the validity of the sequential approach. Therefore, we calculate the ratio of mean-free path $\lambda=(n \sigma)^{-1}$ of the planetesimals, where $n$ is their number density and $\sigma$ the relevant close-encounter cross section, and the close-encounter distance $d_\mathrm{ce}$. Substituting the number density by $\Sigma/2mH$, where $2H$ is the thickness of the planetesimal disc, the mean-free path is given by
\begin{equation}
	\lambda=\frac{2mH}{\sigma\Sigma},	
\end{equation}
where $m$ is the mass of the planetesimal, $H$ is their scale height, and $\Sigma$ is the surface density. The scale height can be expressed in terms of the root-mean square inclination as $H=r\,i_\mathrm{rms}$, the cross section is $\sigma=\pi d_\mathrm{ce}^2$, and we get for the ratio
\begin{align}
	\frac{\lambda}{d_\mathrm{ce}}
	&=\frac{6m_\odot i_\mathrm{rms}}{\pi\Sigma r^2 (d_\mathrm{ce}/R_\mathrm{h})^3}, \nonumber \\
	&=3
	\,\left(\frac{\Sigma}{10\,\mathrm{g}\,\mathrm{cm}^{-2}}\right)^{-1}
	\,\left(\frac{r}{1\,\mathrm{au}}\right)^{-2}
	\,\left(\frac{i_\mathrm{rms}}{5\times10^{-5}}\right)
	\,\left(\frac{d_\mathrm{ce}}{3\,R_\mathrm{h}}\right)^{-3}.
\end{align}

\begin{figure}
	\resizebox{\hsize}{!}{
		\includegraphics{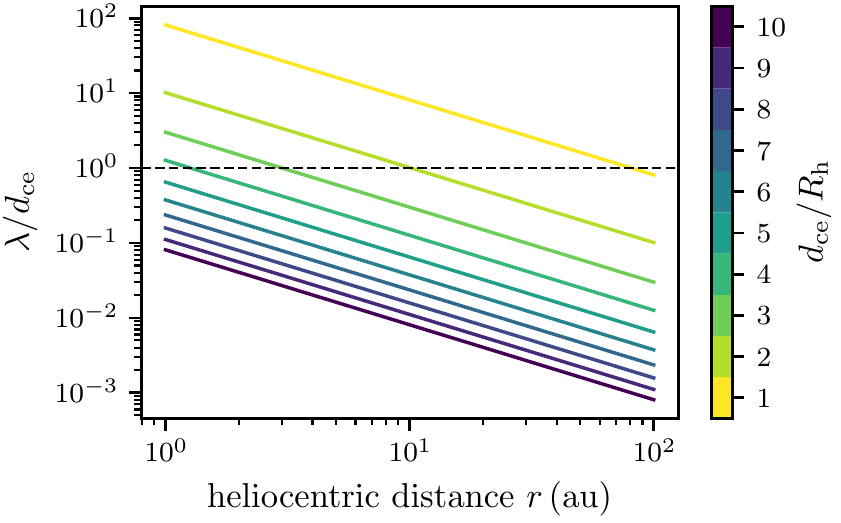}}
	\caption{Ratio of mean-free path and close-encounter distance as function of heliocentric distance. The solid lines represent $\lambda/d_\mathrm{ce}$ for different values of $d_\mathrm{ce}$. For the dashed line, this ratio is unity, indicating where the sequential approach breaks down. A value of $\Sigma_0=10\,\mathrm{g}\,\mathrm{cm}^{-2}$ at $1\,\mathrm{au}$ and $i_\mathrm{rms}=5\times10^{-5}$ was used. Surface density is assumed to be a power-law $\Sigma=\Sigma_0(r/\mathrm{au})^{-1}$}
	\label{fig:paper1_mfp_dce}
\end{figure}
Figure~\ref{fig:paper1_mfp_dce} shows the ratio of mean-free path and close-encounter distance for different values of $d_\mathrm{ce}$ assuming that the surface density varies as $\Sigma=10\,\mathrm{g}\,\mathrm{cm}^{-2}(r/\mathrm{au})^{-1}$. Here, one can see why Fig.~\ref{fig:paper1_dcemax} shows the increased stirring for large $d_\mathrm{ce}$. The ratio $\lambda/d_\mathrm{ce}$ is much lower than unity for $d_\mathrm{ce}\ga5\,R_\mathrm{h}$. This means that in our simulations, most of the planetesimal close-encounter regions, which are defined as the spheres of radius $d_\mathrm{ce}$ around the bodies, start to overlap. In this case, however, integrating the close encounters sequentially will result in stronger stirring. This is because our method evolves the dynamics of the planetesimals as a sequence of kicks without taking the gravitational pull from the other bodies into account resulting in a random-walk-like increase in the eccentricities and inclinations. This can be seen in Fig.~\ref{fig:paper1_dcemax} where at times $\ga3000\,\mathrm{yr}$ and for large $d_\mathrm{ce}$, $e_\mathrm{rms}$ scales with the square-root of time.

\section{Summary}
\label{sec:summary}
The dynamics of planetesimals is important for planet formation because it affects how efficiently they grow to bigger bodies. However, the total number of planetesimals in protoplanetary discs is high and to resolve the dynamics, gravity between all of the bodies would need to be taken into account, which makes it computationally expensive. Therefore, we developed a close-encounter method for simulating a high number of planetesimals. Compared to conventional $N$-body methods which are designed to simulate planetary dynamics with very high accuracy but low $N$, our method focuses on efficiently evolving many interacting small bodies under the assumption that distant encounters can be ignored. For rings of planetesimals this is justified, because significant changes of their orbits only occur during close encounters. With this assumption, the close-encounter method provides the following advantages compared to more detailed $N$-body methods:
\begin{itemize}
	\item Our method is conceptually simple. We employ the same ideas as were developed for symplectic $N$-body integrators. Thanks to this, the numerical integration reduces to solving the Kepler problem and a sequence of few-body problems.
	\item The method can handle planets as well, which makes it applicable for planet formation simulations.
	\item Our method is fast. We achieve a significant speed-up through an optimised close-encounter detection using a cell list (see Sect.~\ref{sec:closeencounterdetection}).
	\item It scales linear when encounters can be integrated sequentially and better than $N^2$ when encounters are grouped together (see Sect.~\ref{sec:computationaltime}).
\end{itemize}
However, there are also limitations to this method: 
\begin{itemize}
	\item The Hamiltonian that is used here describes single-star systems. Therefore, our method is not suitable for simulating binary system. We plan to generalise the method to handle binaries in future work.
	\item The sequential treatment of close encounters is only valid if the close-encounter regions are not overlapping (see Sect.~\ref{sec:discussion}).
	\item If planets are included, their number should remain low. Otherwise our method becomes less efficient, comparable to a conventional $N$-body method, because the code will spend most of the time in calculation interactions between planets (i.e. $\mathcal{H}_\mathrm{B}$). For the ring of equal-mass planetesimals (see Sect.~\ref{sec:equalmassplanetesimals}), the interactions between planets and all other bodies becomes the most expensive process, if we turn $\ga25$ out of the $1000$ planetesimals into planets. For $\ga100$ planets, the interaction part accounts for more than $50\,\%$ of the total computational time. We therefore recommend to limit the number of planets to $\la100$.
	\item The cell list requires a spatial grid which imposes a time step constraint for the code (see Sect.~\ref{sec:gridsizeandtimestep}). The time step is not constant, but depends on the fastest particle in the smallest grid cell.
	\item The method is not symplectic because we do not use a constant time step and because we ignore interactions other than close encounters between planetesimals. Therefore, energy is not conserved which might be a problem for long-term integrations (see Sect.~\ref{sec:energyandangularmomentumconservation}).
\end{itemize}

\begin{acknowledgements}
We thank the anonymous referee for a thorough review that helped to significantly improve the original manuscript. AJ thanks the Swedish Research Council (grant 2018-04867), the Knut and Alice Wallenberg Foundation (grant 2017.0287) and the European Research Council (ERC Consolidator Grant 724687-PLANETESYS) for research support.
\end{acknowledgements}

\bibliographystyle{aa} 
\bibliography{ref} 

\begin{appendix}

\section{Algorithms}
\label{sec:algorithms}
Algorithm~\ref{alg:celllist} describes the algorithm for setting up the list in pseudo-code. The array \texttt{head} of length $n_\mathrm{cell}$ stores for each grid cell the index of the head. The array \texttt{list} of length $N$ and initialised with $0$ contains for each body the index of the next body in that cell.
\begin{algorithm}
	\caption{Construct cell list}
	\label{alg:celllist}
	\begin{algorithmic}[1]
		\Statex
		\State $\texttt{list}\gets0$
		\State $\texttt{head}\gets0$
		\For{$i=1$ to $N$}
		\State calculate $i_\mathrm{cell}$
		\State $\texttt{list}[i] \gets \texttt{head}[i_\mathrm{cell}]$
		\State $\texttt{head}[i_\mathrm{cell}] \gets i$
		\EndFor
	\end{algorithmic}
\end{algorithm}

Algorithm~\ref{alg:detectce} shows how to loop through the cell list. The body pairs that match the criterion for a close encounter are added to a list.
\begin{algorithm}
	\caption{Detect close encounter}
	\label{alg:detectce}
	\begin{algorithmic}[1]
		\Statex
		\For{$i=1$ to $N$}
		\State calculate $i_\mathrm{cell}$ for body $i$
		\For{$j_\mathrm{cell}$ in neighbouring cells}
		\State $j\gets\texttt{head}[j_\mathrm{cell}]$ \Comment{Start with head}
		\While{$j>0$}
		\State check for close encounter between $i$ and $j$
		\If{$d_{ij}<d_\mathrm{ce}$}
		\State add $i,j$ to close-encounter list
		\EndIf
		\State $j \gets \texttt{list}[j]$ \Comment{Continue with next in list}
		\EndWhile
		\EndFor
		\EndFor
	\end{algorithmic}
\end{algorithm}

Algorithm~\ref{alg:groupingce} shows how to group bodies in close-encounter groups. The array \texttt{link} is initialised as $\texttt{link}[i]=i$ for $i=1,\dots,N$. Having the list of close-encounter pairs, a new array of links of length $N$ is created which contains for each body the lowest index of the interacting planetesimals. The indices of the encountering bodies are $i$ and $j$. We set the corresponding value of \texttt{link} via
\begin{align}
	\texttt{link}[i]&=\min(\texttt{link}[i],\texttt{link}[j]), \\
	\texttt{link}[j]&=\min(\texttt{link}[j],\texttt{link}[i])
\end{align}
\citep{Grimm2014}. Bodies that undergo a close encounter are now linked to the body with lowest index in their group. Next, we assign a consecutive group index to the close-encounter groups. We firstly initialise an array \texttt{group} such that $\texttt{group}[i]=i$ for $i=1,\dots,N$ and a counter for the total number of groups $n_\mathrm{groups}$. In a second step, we loop through all bodies, increase the counter if $i=\texttt{link}[i]$, and set the array entry to
\begin{align}
	\texttt{group}[i]=\min(\texttt{group}[\texttt{link}[i]],n_\mathrm{groups}).
\end{align}
This way, we map each body to its group and have a consecutive group index. Finally, we construct a matrix \texttt{cegroup} that contains for each group the indices of the bodies in that group. Another array \texttt{cenumb} contains the total number of bodies within each group.
\begin{algorithm}
	\caption{Grouping of close encounters}
	\label{alg:groupingce}
	\begin{algorithmic}[1]
		\Statex
		\State $\texttt{link}\gets0$
		\For{$i,j$ in close-encounter list}
		\State $\texttt{link}[i]\gets\min(\texttt{link}[i],\texttt{link}[j])$
		\State $\texttt{link}[j]\gets\min(\texttt{link}[j],\texttt{link}[i])$
		\EndFor
		\Statex
		\For{$i=1$ to $N$}
		\State $\texttt{group}[i]\gets i$
		\EndFor
		\Statex
		\State $n_\mathrm{groups}\gets0$
		\For{$i=1$ to $N$}
		\If{$i\mathrel{==}\texttt{link}[i]$}
		\State $n_\mathrm{groups}\gets n_\mathrm{groups}+1$
		\EndIf
		\State\Comment{assign consecutive group index}
		\State $	\texttt{group}[i]\gets\min(\texttt{group}[\texttt{link}[i]],n_\mathrm{groups})$
		\EndFor
		\Statex
		\State $\texttt{cegroup}\gets0$
		\State $\texttt{cenumb}\gets0$
		\For{$i=1$ to $N$}
		\State $j\gets\texttt{group}[i]$
		\State $\texttt{cenumb}[j]\gets \texttt{cenumb}[j]+1$
		\State $k\gets\texttt{cenumb}[j]$
		\State $\texttt{cegroup}[j,k]\gets i$ \Comment{write groups to matrix}
		\EndFor
	\end{algorithmic}
\end{algorithm}

\section{Pebble accretion}
\label{sec:pebbleaccretionprescription}
In this section, we briefly describe our prescription for pebble accretion. Pebble accretion is the process by which a planetesimal grows through the accretion of pebbles \citep{Ormel2010,Lambrechts2012,Johansen2017,Ormel2017}. Pebbles are aerodynamically coupled to and carried along by the gas. When the gravitational pull by the planetesimal is strong enough, pebbles are accreted. This translates into a timescale criterion
\begin{align}
	t_\mathrm{def}=t_\mathrm{fric} \label{eq:accretioncrit}
\end{align}
\citep{Johansen2017,Ormel2017,Lambrechts2019}. Here, $t_\mathrm{def}$ is the deflection timescale and $t_\mathrm{fric}$ is the friction timescale of the pebble.

A pebble approaching the planetesimal at a distance $r_\mathrm{acc}$ with velocity $\Delta v$ is pulled towards the planetesimal on a timescale of
\begin{align}
	t_\mathrm{def}=\frac{\Delta v r_\mathrm{acc}^2}{Gm},
\end{align}
where $m$ is the mass of the planetesimal. The friction timescale is the time it takes for the pebble to adjust to the gas velocity. The relative velocity between pebble and planetesimal, $\Delta v$, can be decomposed into a radial, azimuthal, and vertical component,
\begin{subequations}
\begin{align}
	\Delta v_r &= v_r + \frac{2\eta\tau_\mathrm{f}v_\mathrm{K}}{1+\tau_\mathrm{f}^2}, \\
	\Delta v_\phi &= v_\phi - v_\mathrm{K} + \frac{\eta v_\mathrm{K}}{1+\tau_\mathrm{f}^2} + \frac{3}{2}\Omega_\mathrm{K}r_\mathrm{acc}, \\
	\Delta v_z &= v_z.
\end{align}
\end{subequations}
Here, $\tau_\mathrm{f}=t_\mathrm{fric}\Omega_\mathrm{K}$ is the Stokes number of the pebble and $\eta$ is the pressure gradient of the gas. In the radial direction, the relative velocity is given by the difference between the radial velocity of the planetesimal and the radial drift velocity of the pebble. Ignoring settling, the relative velocity in vertical direction is simply the vertical velocity of the planetesimal. In azimuthal direction, we have to take the difference between the azimuthal velocity of the planetesimal, the azimuthal drift of the pebble, and the Keplerian shear, which is expressed in the third term of $\Delta v_\mathrm{phi}$. By using this expression for $\Delta v$, eccentricity and inclination of the planetesimal is taken into account and we can write
\begin{align}
	\Delta v = \sqrt{\Delta v_r^2 + \Delta v_\phi^2 + \Delta v_z^2}	
\end{align}
\citep{Lambrechts2019}.

Because $\Delta v$ depends on the accretion radius, we solve the criterion Eq.~\ref{eq:accretioncrit} iteratively to obtain the accretion radius \citep{Lambrechts2019}. The accretion radius according to Eq.~\ref{eq:accretioncrit} is obtained with the assumption that the encounter timescale is long enough for the pebble to settle towards the planetesimals due to the action of gas drag. Therefore, we have to check whether this is the case by comparing the encounter timescale to the friction timescale of the pebble \citep{Ormel2017}. The encounter timescale, $t_\mathrm{enc}$, is the time it takes for the pebble to pass the gravitational reach of the planetesimal and can be expressed as
\begin{align}
	t_\mathrm{enc}=\frac{2r_\mathrm{acc}}{\Delta v}
\end{align}
\citep{Ormel2017}. We calculate $t_\mathrm{enc}$ and compare it to $t_\mathrm{fric}$. If $t_\mathrm{enc}<t_\mathrm{fric}$, we set the accretion radius to
\begin{align}
	r_\mathrm{acc}=R\sqrt{1+\left(\frac{v_\mathrm{esc}}{\Delta v}\right)^2},
\end{align}
which is the accretion radius for gravitational focusing. $v_\mathrm{esc}=\sqrt{2Gm/R}$ is the escape velocity for a body of radius $R$ \citep{Ormel2010,Ormel2017}.

With the accretion radius at hand, we can determine the mass accretion rate. If the accretion radius is larger than the scale height of the pebble layer, $H_\mathrm{peb}$, accretion takes place in the 2D regime and the mass accretion rate becomes
\begin{align}
	\dot{m}_{2\mathrm{D}}=2\pi r_\mathrm{acc} \Delta v \Sigma_\mathrm{peb}.
\end{align}
If, on the other, $r_\mathrm{acc}<H_\mathrm{peb}$ accretion is in 3D and the mass accretion rate reads
\begin{align}
	\dot{m}_\mathrm{3\mathrm{D}}=\frac{\Sigma_\mathrm{peb}}{\sqrt{2\pi}H_\mathrm{peb}}\pi r_\mathrm{acc}^2 \Delta v
\end{align}
\citep{Lambrechts2012,Johansen2017,Ormel2017,Lambrechts2019}. Typically, planetesimals start in the 3D regime and make a transition into 2D when they become massive enough. Accretion of pebbles stops when the body reaches its pebble isolation mass. In that case, the body opens a gap in the gas which creates a pressure bump and prevents pebbles from drifting in. We take the pebble isolation mass as
\begin{align}
	m_\mathrm{iso}\approx 20\left(\frac{H_\mathrm{g}/r}{0.05}\right)^3 M_\oplus,
\end{align}
where $H/r$ is the aspect ratio of the gas disc \citep{Johansen2017}. Accretion also stops when the inclined planetesimal reaches a height that is above the pebble scale-height. And lastly, we take mutual filtering into account. That means that the pebble flux decreases towards inner orbits because every planetesimal on an outer orbit takes out a certain portion of the pebble flux \citep{Liu2019}.

\begin{figure}
	\resizebox{\hsize}{!}{
		\includegraphics{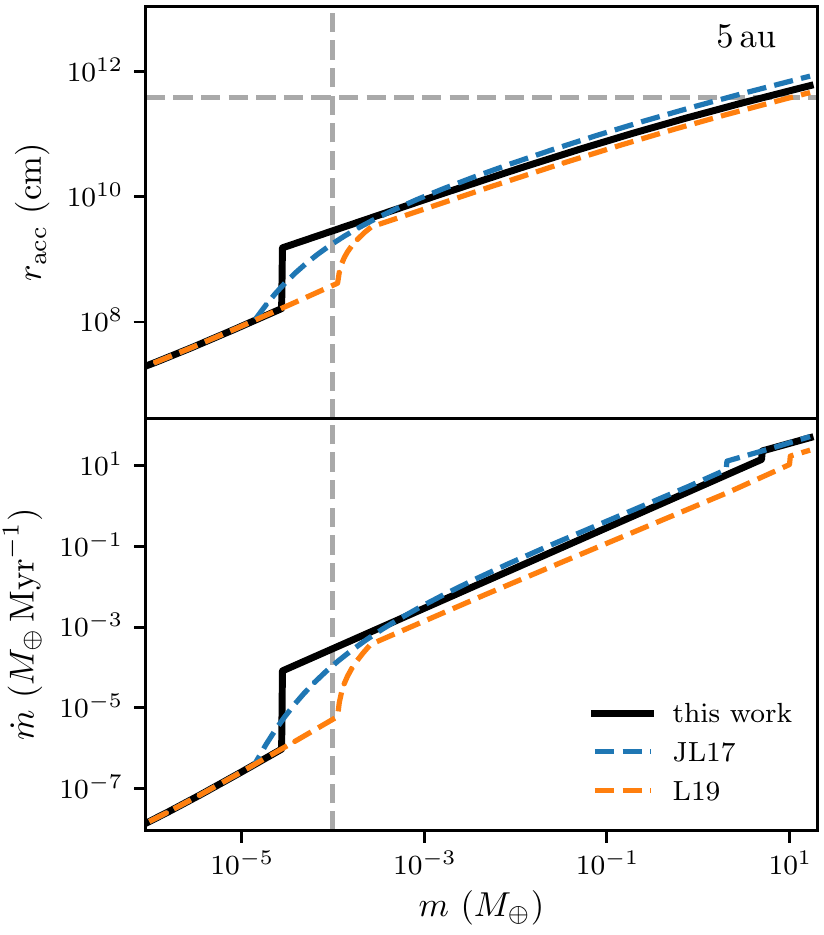}}
	\caption{Accretion radius and mass accretion rate for pebble accretion at $5\,\mathrm{au}$. The figure shows $r_\mathrm{acc}$ (\textit{top panel}) and $\dot{m}$ (\textit{bottom panel}) as a function of planetesimal mass. We compare our prescription (black solid line) to the models developed in \citet{Johansen2017} (JL17, blue dashed line) and \citet{Lambrechts2019} (L19, dashed orange line). The vertical dashed line indicates the initial mass of the planetesimals used in Sect.~\ref{sec:pebbleaccretion}. The horizontal dashed line (\textit{top panel}) is the pebble scale-height marks the transition from 3D to 2D accretion.}
	\label{fig:paper1_pamodel5au}
\end{figure}
\begin{figure}
	\resizebox{\hsize}{!}{
		\includegraphics{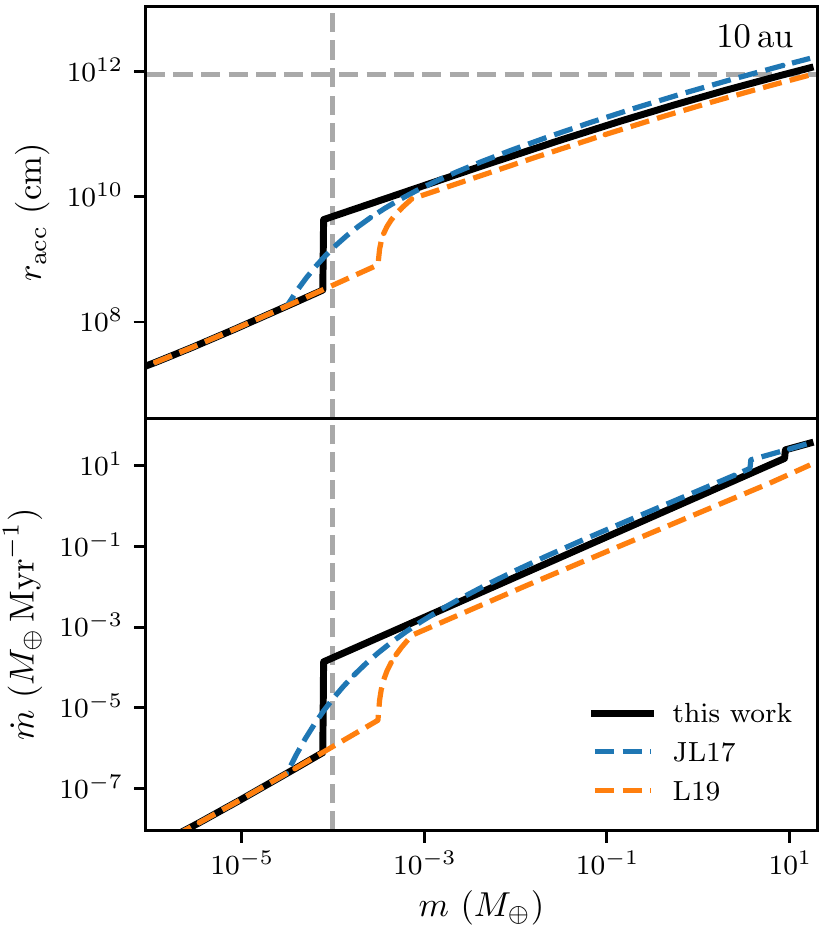}}
	\caption{Accretion radius and mass accretion rate for pebble accretion at $10\,\mathrm{au}$. The figure shows $r_\mathrm{acc}$ (\textit{top panel}) and $\dot{m}$ (\textit{bottom panel}) as a function of planetesimal mass. We compare our prescription (black solid line) to the models developed in \citet{Johansen2017} (JL17, blue dashed line) and \citet{Lambrechts2019} (L19, dashed orange line). The vertical dashed line indicates the initial mass of the planetesimals used in Sect.~\ref{sec:pebbleaccretion}. The horizontal dashed line (\textit{top panel}) is the pebble scale-height marks the transition from 3D to 2D accretion.}
	\label{fig:paper1_pamodel10au}
\end{figure}
Figures~\ref{fig:paper1_pamodel5au} and \ref{fig:paper1_pamodel10au} show the accretion radius and mass accretion rate for pebble accretion at $5\,\mathrm{au}$ and $10\,\mathrm{au}$, respectively. These represent the inner and outer edge of the ring of planetesimals simulated in Sect.~\ref{sec:pebbleaccretion}. We furthermore used the same parameters for the Stokes number, the pebble flux, and the ratio of pebble to gas scale-height as in our simulation to generate these plots. Our prescription is in general very similar to \citet{Johansen2017} and \citet{Lambrechts2019}. We have some differences because we do not include the weak-coupling Bondi regime \citep{Ormel2010,Lambrechts2019} and our mass accretion rates are slightly higher than in \citet{Lambrechts2019}. The vertical dashed lines in Figs.~\ref{fig:paper1_pamodel5au} and \ref{fig:paper1_pamodel10au} indicate the initial mass of the planetesimals ($10^{-4}\,M_\oplus$) used in our simulation. The biggest differences compared to the other two models occur around that mass for heliocentric distances between $5\,\mathrm{au}$ and $10\,\mathrm{au}$. We therefore overestimate the growth of the low-mass planetesimals, but capture the embryos fairly well. The transition from 3D to 2D accretion is indicated by the horizontal dashed line, which shows the pebble scale-height. Only for the most massive bodies, $\ga5\,M_\oplus$, accretion is in the 2D regime. The prescription used here is thus sufficient for testing purposes without producing entirely wrong results.

\section{Gas drag and migration}
\label{sec:gasdragandmigration}

\subsection{Gas drag}
\label{sec:gasdrag}
We implement gas drag through the protoplanetary disc acting on the planetesimals. We use as an example here the simple Minimum Mass Solar Nebula (MMSN) model \citep{Weidenschilling1977b,Hayashi1981}. Surface density and temperature of the gas decrease with increasing heliocentric distance as power laws with respective slopes $\alpha$ and $\beta$,
\begin{align}
	\Sigma&=1700\,\mathrm{g}\,\mathrm{cm}^{-2}\,\left(\frac{r}{\mathrm{au}}\right)^{-\alpha}, \\
	T&=280\,\mathrm{K}\,\left(\frac{r}{\mathrm{au}}\right)^{-\beta}.
\end{align}
The drag force acting on the planetesimal is determined by a drag coefficient $C_\mathrm{D}$, the gas density $\rho$, the aerodynamic cross-section of the body of radius $R$, and the relative velocity $\vect{v}_\mathrm{rel}=\vect{v}-\vect{v}_\mathrm{g}$ between the planetesimal and the gas:
\begin{align}
	\vect{F}_\mathrm{D}=-\frac{1}{2}C_\mathrm{D}\rho\pi R^2|\vect{v}_\mathrm{rel}|\vect{v}_\mathrm{rel}.
\end{align}
We assume the disc to be in vertical hydrostatic equilibrium. Therefore, the density is
\begin{align}
	\rho=\frac{\Sigma}{\sqrt{2\pi}H_\mathrm{g}}
	\exp{\left(-\frac{z^2}{2H_\mathrm{g}^2}\right)},
\end{align}
where $H_\mathrm{g}=c_\mathrm{s}/\Omega_\mathrm{K}$ is the gas scale-height of the disc, which is the ratio of sound speed $c_\mathrm{s}$ and Keplerian angular frequency $\Omega_\mathrm{K}$. The gas velocity is sub-Keplerian and can be written as
\begin{align}
	v_\mathrm{g}=v_\mathrm{K}(1-\eta),
\end{align}
where $\eta$ is set by the pressure gradient
\begin{align}
	\eta = -\frac{1}{2}\left(\frac{c_\mathrm{s}}{v_\mathrm{K}}\right)^2\frac{\partial\log P}{\partial\log r}.
\end{align}
The drag coefficient $C_\mathrm{D}$ depends on both the properties of the planetesimal and the disc gas. If the size of the body is smaller than the mean-free path of the gas $\lambda_\mathrm{g}$, that is $R/\lambda_\mathrm{g}<4/9$, the body is in the Epstein drag regime and $C_\mathrm{d}$ becomes \citep{Epstein1924}
\begin{align}
	C_\mathrm{d}=\frac{8}{3}\frac{v_\mathrm{thm}}{|\vect{v}_\mathrm{rel}|},
\end{align}
where $v_\mathrm{thm}=\sqrt{8/\pi}c_\mathrm{s}$ is the thermal velocity of the gas. On the other hand, if the body is larger, the drag coefficient is more complicated because the drag arises from the hydrodynamic flow across the object. Depending on the Reynolds number $Re=2R|\vect{v}_\mathrm{rel}|/\nu$, where $\nu=(1/3)\lambda_\mathrm{g}v_\mathrm{thm}$ is the molecular viscosity, the drag coefficient becomes
\begin{align}
	C_\mathrm{d}=
	\begin{cases}
		24\,Re^{-1}& Re<1 \\
		24\,Re^{-0.6}& 1\le Re<800 \\
		0.44& Re\ge 800
	\end{cases}
\end{align}
\citep{Whipple1972,Weidenschilling1977}. If the body moves supersonically through the gas, that is $|\vect{v}_\mathrm{rel}|>c_\mathrm{s}$, the drag coefficient takes on a constant value of $C_\mathrm{d}=2$ \citep{Brasser2007}.

\subsection{Type-I migration}
\label{sec:type1migration}
In addition to gas drag, we also implement type-I migration and the damping of eccentricity and inclination by adding a force which mimics the results of hydrodynamic simulations \citep{Cresswell2008}. The fundamental damping timescale is
\begin{align}
	t_\mathrm{wave}=\frac{M_\odot}{M}\frac{M_\odot}{\Sigma r^2}\left(\frac{H_\mathrm{g}}{r}\right)^4\Omega_\mathrm{K}^{-1}
\end{align}
which is the timescale for damping of semi-major axis $a$, eccentricity $e$, and inclination $i$ due to the excitation of density waves in the disc by the planetesimal \citep{Tanaka2004}.

\citet{Cresswell2008} then provide timescales for the damping of $a$, $e$, and $i$ that fit their hydrodynamical simulations most accurately. These are
\begin{align}
	t_e &= \frac{t_\mathrm{wave}}{0.780}\left[1-0.14\,\tilde{e}^2+0.06\,\tilde{e}^3+0.18\,\tilde{e}\,\tilde{i}^2\right], \\
	t_i &= \frac{t_\mathrm{wave}}{0.544}\left[1-0.30\,\tilde{i}^2+0.24\,\tilde{i}^3+0.14\,\tilde{e}^2\,\tilde{i}\right], \\
	t_a &= \frac{2t_\mathrm{wave}}{2.7+1.1\,\alpha}\left(\frac{H_\mathrm{g}}{r}\right)^{-2}\left\{P(\tilde{e})+\frac{P(\tilde{e})}{|P(\tilde{e})|}\right. \nonumber \\
	&\times\left.\left[0.070\,\tilde{i}+0.085\,\tilde{i}^4-0.080\,\tilde{e}\,\tilde{i}^2\right]\right\},
\end{align}
with
\begin{align}
	P(\tilde{e})=\frac{1+(\tilde{e}/2.25)^{1.2}+(\tilde{e}/2.84)^6}{1-(\tilde{e}/2.02)^4}.
\end{align}
$\alpha$ is the surface density slope of the gas and eccentricity and inclination are divided by the aspect ratio of the disc as $\tilde{e}=e/(H_\mathrm{g}/r)$ and	$\tilde{i}=i/(H_\mathrm{g}/r)$. The accelerations that act on the planetesimal are then
\begin{align}
	\vect{a}_a &= -\frac{\vect{v}}{t_a}, \\
	\vect{a}_e &= -2\frac{(\vect{v}\cdot\vect{r})\vect{r}}{r^2t_e}, \\
	\vect{a}_i &= -\frac{v_z}{t_i}\vect{e}_z,
\end{align}
where $\vect{e}_z$ is the unit vector in $z$-direction.

\end{appendix}

\end{document}